\documentclass[12pt,preprint]{aastex}

\shorttitle{}
\shortauthors{Nesvorn\'y et al.}

\begin{document}
\baselineskip 19.pt

\title{Observed Binary Fraction Sets Limits on the Extent \\ 
of Collisional Grinding in the Kuiper Belt}

\author{David Nesvorn\'y$^1$, David Vokrouhlick\'y$^{1,2}$, William F. Bottke$^1$,\\ 
Keith Noll$^{3}$, Harold F. Levison$^1$}
\affil{(1) Department of Space Studies, Southwest Research Institute,\\
1050 Walnut St., Suite 300, Boulder, CO 80302, USA}
\affil{(2) Institute of Astronomy, Charles University, \\
V Hole\v{s}ovi\v{c}k\'ach 2, CZ-18000, Prague 8, Czech Republic}
\affil{(3) Space Telescope Science Institute, 3700 San Martin Dr., Baltimore, 
MD 21218, USA}

\begin{abstract}
The size distribution in the cold classical Kuiper belt can be approximated by two idealized 
power laws: one with steep slope for radii $R>R^*$ and one with shallow slope for $R<R^*$, where 
$R^*\sim25$-50 km. Previous works suggested that the SFD roll-over at $R^*$ can be the result 
of extensive collisional grinding in the Kuiper belt that led to the catastrophic disruption 
of most bodies with $R<R^*$. Here we use a new code to test the effect of collisions in the 
Kuiper belt. We find that the observed roll-over could indeed be explained by collisional grinding 
provided that the initial mass in large bodies was much larger than the one in the present 
Kuiper belt, and was dynamically depleted. In addition to the size distribution changes, 
our code also tracks the effects of collisions on binary systems.  We find that it is generally 
easier to dissolve wide binary systems, such as the ones existing in the cold Kuiper belt today, 
than to catastrophically disrupt objects with $R\sim R^*$. Thus, the binary survival sets 
important limits on the extent of collisional grinding in the Kuiper belt. We find that the 
extensive collisional grinding required to produce the SFD roll-over at $R^*$ would imply a 
strong gradient of the binary fraction with $R$ and separation, because it is generally easier to 
dissolve binaries with small components and/or those with wide orbits. The expected binary 
fraction for $R\lesssim R^*$ is $\lesssim$0.1. The present observational data do not show such 
a gradient. Instead, they suggest a large binary fraction of $\sim$0.4 for $R=30$-40 km. This 
may indicate that the roll-over was {\it not} produced by disruptive collisions, but is instead
a fossil remnant of the KBO formation process. 
\end{abstract}


\section{Introduction}

The cold Classical Kuiper Belt, hereafter cold CKB, is a population of trans-Neptunian 
bodies dynamically defined as having orbits with semimajor axis $a=42$-48 AU, perihelion distances 
that are large enough to avoid close encounters to Neptune, and low inclinations ($i\lesssim5^\circ$). 

Cold CKB objects (cold CKBOs) show several properties that distinguish them from other 
populations in the trans-Neptunian region. Specifically, cold CKBOs have 
distinctly red colors (Tegler \& Romanishin 2000) that may have resulted from space weathering
of surface ices, such as methanol (e.g., Schaller 2010), that are stable beyond 30~AU. A large fraction 
of 100-km-class cold CKBOs are binaries with wide separations and similar size components (Noll et 
al. 2008a,b). These binaries are practically absent in the dynamically hot, resonant and scattered 
populations. The albedos of cold CKBOs are generally higher than those of the dynamically hot CKBOs 
(Brucker et al. 2009). And finally, the magnitude distribution of cold CKBOs is markedly different 
from those of the hot and scattered populations, in that it shows a steep slope for the large objects: 
$\Sigma(m_{\rm R}){\rm d}m_{\rm R} \sim 10^{\alpha m_{\rm R}}$ with $\alpha\sim0.82$, where 
$\Sigma(m_{\rm R})$ is the number of objects per square degree with R magnitude $m_{\rm R}$ 
(Fraser et al. 2010). 

The magnitude distribution of CKBOs shows a roll-over to a shallower slope at magnitudes 
$m_{\rm R}^* \sim 25$ (Bernstein et al. 2004, see Petit et al. 2008 for a review). This feature is thought 
to occur due to a genuine change in the slope of the size frequency distribution (SFD), 
because extreme albedo variations would need to be invoked if the SFD slope were constant over $m_{\rm R}^*$. 
Using albedo $p_V=0.2$ for the cold CKBOs (Brucker et al. 2009) and $m_{\rm R}^* = 25$ (Fuentes et al. 
2009), the roll-over radius is $R^* = 37$ km. Allowing for uncertainties in $p_V$ and $m_{\rm R}^*$, 
it is probably safe to assume that $25 \lesssim R^* \lesssim 50$ km. 

For $m_{\rm R}<m_{\rm R}^*$, $\alpha\sim0.82$ from Fraser et al. (2010) implies $N(R){\rm d}R \sim R^{-q}{\rm d}R$ 
with $q=q_>\sim5$ for $R>R^*$ ($q=5\alpha+1$). For $m_{\rm R}>m_{\rm R}^*$, $\alpha\sim0.2$-0.3 from 
Fuentes \& Holman (2008) and Fuentes et al. 
(2009) implies that $q=q_<\sim2$-3 for $R<R^*$. These SFD slope estimates will be used to constrain
our model (\S3). Note that the shallow SFD slope for $m_{\rm R}>m_{\rm R}^*$ is an empirical approximation 
that fits observations only up to the limiting magnitude $m_{\rm R}\sim27$ (Fuentes et al. 2009) or 
equivalently down to $R\sim15$-20 km, depending on albedo. The actual SFD can be wavy, just like
the SFD of the asteroid belt at these radii (e.g., Bottke et al. 2005). Indeed, the occultation data
from Schlichting et al. (2009) may indicate that the number of $R>250$ m KBOs is significantly 
larger than what would be expected from extrapolating the SFD with $q \lesssim 3$ from $R^*$ down to 
$R=250$ m.

The SFD roll-over at $R^*$ may be telling us something important about the formation and evolution 
of cold CKB. As first suggested by Pan \& Sari (2005, hereafter PS05), the shallow SFD slope
for $R<R^*$ can be the result of extensive collisional grinding in the cold CKB. Indeed, the SFD slope index 
below $R^*$, $q_<\sim2$-3, is similar to the slope index expected for a collisionally evolved 
population that reaches the equilibrium slope (Dohnanyi 1969, O'Brien \& 
Greenberg 2003).\footnote{Dohnanyi's equilibrium slope $q=3.5$ (Dohnanyi 1969) applies to a situation 
in which the object's strength is independent of its size. In the gravity regime of impacts, applicable 
the size range relevant here, the strength increases with size and $q\sim3$ (O'Brien \& Greenberg 2003).} 

PS05 carried out order-of-magnitude analytic calculations to support their argument. They used 
an idealized two-slope SFD with $q_<\approx3$ and $q_>=5$, and estimated $R^*$ by postulating that bodies 
of radii $R \leq R^*$ are disrupted over 4.5~Gy. As presented, their results were meant to 
apply to the case in which the number of bodies with $R>R^*$ has not changed over 4.5~Gy, and was 
equal to the number of bodies in the present KB (but see below). They found that $R^*= 20$-50 km, 
in agreement with observations, with the exact value of $R^*$ mainly depending on the assumed disruption 
law.

We were unable to reproduce these analytical results with the numerical code described in \S2. 
Specifically, using PS05's assumptions, we always obtained the break well below 10~km (see \S3.1). 
Since our code was thoroughly tested, and gives results that are essentially identical to those of other 
numerical codes (Weidenschilling et al. 1997, Stern \& Colwell 1997, Kenyon \& Bromley 2001, 
Bottke et al. 2005, Fraser 2009), one might wonder why it fails so badly in reproducing the PS05 results. 
While some unspecified numerical problems, common to all tested codes, can be responsible, there are 
also a few issues with the PS05 calculations that need to be clarified.

PS05 claim that they normalize the population at $R^*$ to be equivalent to KB
with $4\times10^4$ bodies larger than 100 km. A thorough reading
of the description of their normalization procedure reveals, however, that they mean $R>100$~km rather 
than diameter $D>100$ km. Using this normalization, they therefore have $4\times10^4$ bodies with 
$R>100$ km, which implies $\sim$$6.4\times10^5$ bodies with $R>50$~km, if $R^*<50$ km and $q_>=5$. 
For comparison, observations indicate that there should only be 30,000 to 50,000 KBOs with $R>50$ km 
(Jewitt \& Luu 1995, Trujillo et al. 2001, Fuentes \& Holman 2008). 
PS05 therefore effectively assume a population of $R>R^*$ bodies that is a factor of $\sim$16 larger 
than the one existing in the KB today. Since, according to PS05's eq. (8), $R^*$ scales with 
$~f^{1/3}$, where $f=16$ is the augmentation factor, we find that $R^*$ should be about a factor of 
2.5 lower than what was claimed in PS05, or $R^*=8$-20 km, if $f=1$. 

\begin{figure}
\epsscale{0.6}
\plotone{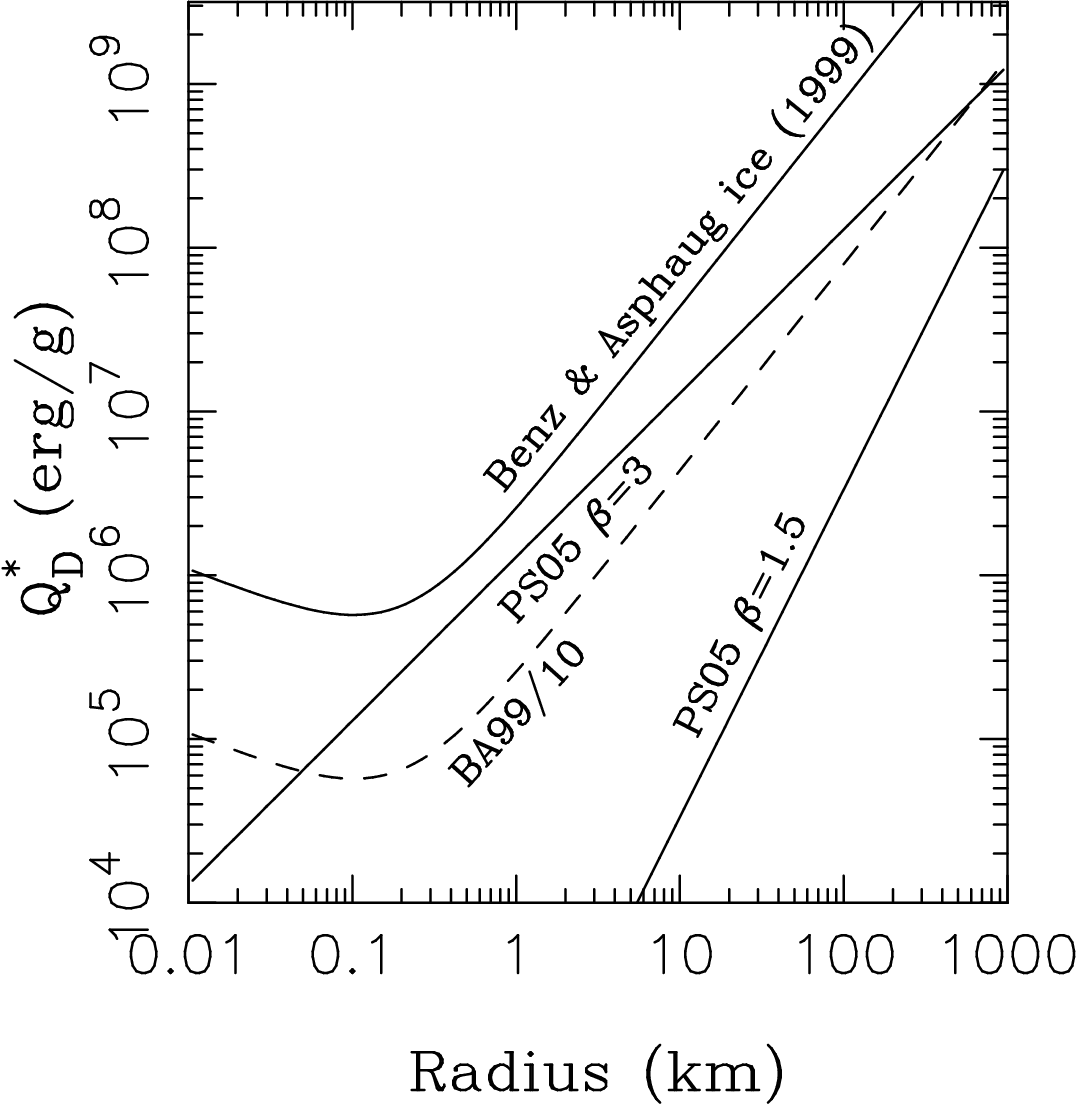}
\caption{Contrasting assumptions on the strength of KBOs. The plot shows the specific energy 
of catastrophic disruption, $Q^*_{\rm D}$, as a function of radius of the impacted body.
The upper solid curve shows $Q^*_{\rm D}$ as determined from hydrodynamic simulations of 
impacts on water ice (Benz \& Asphaug 1999). The dashed curve shows Benz \& Asphaug's ice $Q^*_{\rm D}$ 
divided by 10, which is the weakest disruption law considered in our work, and also roughly 
the smallest $Q^*_{\rm D}$ value in the gravity regime found in numerical simulations of 
impacts into porous rubble piles (Leinhardt \& Stewart 2008, Stewart \& Leinhardt 2008).   
The two solid straight lines show $Q^*_{\rm D}$ that we computed from the strong ($\beta=3$)
and weak ($\beta=3/2$) disruption laws of PS05.}
\label{qstar}
\end{figure}

The second issue is related to the PS05's disruption laws. To obtain large values of $R^*$, they
assumed extremely weak disruption laws ($\beta=1.5$ in the notation of their eq. 5), 
where the specific impact energy needed to catastrophically disrupt and disperse an object of size 
$R$, $Q^*_D(R)$, was up to $\sim10^3$ times lower than the one estimated for water ice by Benz \& 
Asphaug (1999) (Fig. \ref{qstar}). Such extremely weak disruption laws are probably unrealistic. 
More realistic disruption laws, such as those advocated by Stewart \& Leinhardt (2009) and Leinhardt 
\& Stewart (2009), are similar to the strongest disruption law considered by PS05 ($\beta = 3$ 
in their eq. 5). With these laws, $R^*<10$ km according to PS05 (for $f=1$), which is more in line with 
the results of our numerical code. We therefore find that either the number of large objects was much 
larger than the one in the present KB, implying faster initial grinding, or the observed roll-over 
at $R^*=25$-50 km must be due to something else.

The SFD roll-over can be a signature of the accretional processes that were active during the KBO
formation (see Kenyon et al. 2008 and Chiang \& Youdin 2010 for reviews). 
Interestingly, the asteroid belt, thought to have formed by similar processes, also 
shows a roll-over at $R\approx50$ km. Given better constraints that we have on the fragmentation 
processes in the asteroid belt (e.g., survival of the Vesta's crust, number of asteroid 
families, etc.), it has been established that the roll-over was {\it not} produced by collisional 
grinding (Bottke et al. 2005). Instead, the asteroid SFD roll-over was likely created by 
accretional processes, and constrains them in important ways (Morbidelli et al. 2009). The 
distinction between fragmentation and accretion signatures is more difficult to make in the
Kuiper belt, where we are lacking good constraints.

The binary KBOs can provide an interesting constraint on the amount of collisional grinding in 
the Kuiper belt. Recent observations indicate that $\sim$30\% of 100-km-class cold CKBOs are 
binaries (Noll et al. 2008a,b; $>$0.06 arcsec separation, $<$2 mag magnitude contrast). 
The properties of known binary KBOs differ markedly from those of the main-belt and near-Earth 
asteroid binaries (Merline et al. 2002, Noll et al. 2008a). The 100-km-class binary KBOs 
identified so far are widely separated and their components are similar in size. These properties 
defy standard ideas about processes of binary formation involving collisional and rotational 
disruption, debris re-accretion, and tidal evolution of satellite orbits (Stevenson et al. 1986). 
They suggest that most binary KBOs are remnants from the earliest days of the Solar System. 
Indeed, all models developed so far postulate that binary formation was contemporary to the 
formation of KBOs themselves (Weidenschilling 2002, Goldreich et al. 2002, Funato et al. 2004, 
Astakhov et al. 2005, Nesvorn\'y et al. 2010).

The survival of binary KBOs after their formation is an open problem. Petit \& Mousis (2004) 
estimated that several known binary KBOs, such as 1998 WW$_{31}$, 2001 QW$_{322}$ and 2000 CF$_{105}$, 
have lifetimes against collisional unbinding that are much shorter than the age of the solar 
system. These estimates were based on the steep SFD adopted by Petit \& Mousis that was
extended down to $R=5$ km. This assumption favors binary disruption, because of the large 
number of available impactors. When we update 
Petit \& Mousis' work with the more recent estimates of the SFD in the Kuiper belt (e.g.,
$R^*=25$-50 km for cold CKBOs; see discussion above), we find that a typical 100-km-class wide 
binary CKBO is unlikely to be disrupted over the age of the solar system ($\lesssim1$\% 
probability), except if the KB (or its source population) was more massive/erosive in the past. 

Here we use the observed binary fraction in the cold CKB to determine how it limits the amount 
of collisional grinding in the Kuiper belt. We make different assumptions on the initial state 
and history of the cold CKB, and the related populations, and identify cases that lead to the
SFD break of cold CKBOs at $R^*=25$-50 km. We then evaluate the survival of binary CKBOs in each 
of these scenarios. If the roll-over at $R^*=25$-50 km was produced by collisional grinding, 
we find that the binary fraction should show a strong gradient with radius and binary 
separation, because it is generally easier to dissolve binaries with small components or those with 
very wide orbits (\S3). The absence of such a gradient would indicate that the roll-over is 
accretional. 

\section{Modeling Collisional Evolution and Binary Survival}

\subsection{Collisional Evolution Code}

Our collisional modeling simulations employ {\it Boulder}, a new code capable of simulating the 
collisional fragmentation of multiple planetesimal populations using a statistical particle-in-the-box 
approach. It was constructed along the lines of other published codes (Weidenschilling et al. 1997,
Kenyon \& Bromley 2001). A full description of the {\it Boulder} code, how it was tested, and its 
application to both accretion and collisional evolution of the early asteroid belt, can be found 
in Morbidelli et al. (2009). Examples of its previous use for the asteroid belt, Hildas, Trojans, 
irregular satellites, and primordial trans-planetary disk are described in Levison et al. (2009) 
and Bottke et al. (2010). 

The code's procedure for modeling impacts is as follows. For a given impact between a projectile 
and a target object, the code computes the specific impact energy $Q$, defined as the kinetic energy of 
the projectile divided by the target mass, and the critical impact energy $Q^*_D$, defined as the 
energy per unit target mass needed to disrupt and disperse 50\% of the target (e.g., Davis et al. 
2002). For reference, $Q < Q^*_D$ values correspond to cratering events, $Q \approx Q^*_D$ correspond 
to barely catastrophic disruption events, and $Q > Q^*_D$ correspond to super-catastrophic 
disruption events.

For each collision identified by the code, the mass of the largest remnant is computed from the 
scaling laws found in hydrodynamic simulations of impacts (Benz \& Asphaug 1999, Leinhardt \& 
Stewart 2009, Stewart \& Leinhardt 2009). The mass of the largest fragment and the slope of the 
power-law SFD produced by each collision is set as a function of $Q/Q^*_D$ by empirical fits 
to the hydrocode results of Durda et al. (2004, 2007) and Nesvorn\'y et al. (2006). See Bottke et al. 
(2010) for the explicit definition of these fits. These results apply to monolithic target bodies. 
We also tested approximate scaling laws for impacts on pre-fragmented and rubble-pile targets. Again, 
these scaling relations were drawn from the fits to hydrocode impact simulations (Benavidez et al. 
2011). 

The $Q^*_D$ function was assumed to split the difference between the impact experiments of Benz \& 
Asphaug (1999), who used a strong formulation for ice, and those of Leinhardt \& Stewart (2009), 
who used the finite volume shock physics code to perform simulations into what they describe 
as weak ice. To do this, we divide the Benz \& Asphaug (1999) strong-ice $Q^*_D$ function by a 
factor, $f_Q$. We typically used $f_Q = 3$, 5, and 10. Note that because we sampled a broad section of 
parameter space, we chose not to include still more complicating factors (e.g., $Q^*_D$ may 
vary with impact velocity, etc.). The bulk density was set to $\rho=1$~g~cm$^{-3}$.
 
The main input parameters for the {\it Boulder} code are: the (i) initial SFD of the simulated populations 
(see \S3); (ii) intrinsic collision probability, $P_i$, defined as the probability that a single 
member of the impacting population will hit a unit area of a body in the target population over a 
unit of time, and (iii) mean impact speed, $v_i$. For collisions between present-day CKBOs, we 
used $P_i=4 \times 10^{-22}$ km$^{-2}$ yr$^{-1}$ and $v_i=1$ km s$^{-1}$ (Davis \& Farinella 1997,
Dell'Oro et al. 2001).\footnote{Note that, 
according to PS05, $P_i$ can be approximated by $\sim\Omega/A$, where $\Omega$ is the typical 
orbital angular velocity in the Kuiper belt, and $A$ is the Kuiper belt's area in the plane 
of the solar system. With $\Omega=0.022$~yr$^{-1}$ and $A=1200$ AU$^2$, PS05 thus effectively 
have $P_i=10^{-21}$ km$^{-2}$ yr$^{-1}$, a value larger by a factor of 2.5 than the one adopted in 
this work.} See \S3 for our specific choices of $P_i$ and $v_i$ for other populations.  

\subsection{Binary Survival Code}

Petit \& Mousis (2004) identified three main processes that can dissolve KBO binaries:
(1) One of the component is hit by a small impactor. While the component survives essentially 
intact, except for a new crater on its surface, the linear momentum transferred from the 
impactor imparts a `kick' on the velocity vector of the binary orbit. If the kick is large 
enough, the component becomes unbound from its companion.
(2) One of the components can be shattered by a large impactor. The fragments produced the 
object's breakup are expected to escape on unbound trajectories because the ejection speeds ($\sim$100 m 
s$^{-1}$) largely exceed that of the binary orbit ($\sim$1 m s$^{-1}$). (3) The binary system has a close 
gravitational encounter with another KBO. The tidal gravity of that object can unbind the 
binary provided that the object is massive enough (e.g., Stern et al. 2003). 

Petit \& Mousis (2004) found that mechanism (1) is by far the most efficient way to dissolve
binaries in the present KB. Thus, we focus on modeling (1) in this work. Mechanism (2) 
is also considered, but we find that its effects are negligible compared to (1), except if extremely 
weak (and probably unrealistic) disruption laws are adopted. This result stems from the fact that 
it is generally easier to dissolve a wide KBO binary orbit by an impact than to physically 
disrupt a 100-km object. Mechanism (3) could have been important during the early phases of 
the Kuiper belt evolution provided that the cold CKB overlapped with a population of 
numerous, very massive KBOs. We do not model (3) in this work because we do not yet have a 
detailed understanding of these early stages.

We will assume in the following that small impacts with $Q \ll Q^*_D$ can be treated as inelastic 
collisions. Thus, we will ignore any linear momentum that can be potentially carried away by ejected 
fragments. In this approximation, and assuming that $m_i \ll m$, where 
$m_i$ and $m$ are the impactor and binary component masses, the 
velocity vector of the binary orbit, ${\bf v}$, will change by  
\begin{equation}
 \delta {\bf v} \simeq \frac{m_i}{m}\,{\bf v}_i \; . 
\label{dv}
\end{equation} 
See Dell'Oro \& Cellino (2007) for a discussion of the linear momentum transfer for different 
impact angles, and in the case where the escaping ejecta affect the linear momentum budget.

The binary components are assumed to have equal mass, which should be a good approximation for the
real binaries in the cold CKB (see \S1). This assumption is conservative in the sense that it is generally 
harder to disrupt a binary with equal-size components than the one in which the secondary is smaller than 
the primary (if other parameters are the same). By using equal-size binaries we may thus slightly 
underestimate the real decay rate.

The radial, tangential and normal components of $\delta {\bf v}$ will be denoted by 
$\delta v_R$, $\delta v_T$ and $\delta v_Z$, respectively, in the following. With this notation, the 
semimajor axis $a$, eccentricity $e$ and inclination $I$ of the binary orbit will change by 
\begin{eqnarray}
 \delta a &=& \frac{2}{n\eta} \left[\delta v_T \left(1+e\cos f\right)
  + \delta v_R\, e \sin f\right]\; , \label{da} \\
 \delta e &=& \frac{\eta}{na} \left[\delta v_R \sin f
  + \delta v_T \left(\cos f+\frac{e+\cos f}{1+e\cos f}\right)\right]\; , \\
 \delta (\cos I) &=& \frac{\eta}{na}\, \delta v_Z\, \frac{\sin(\omega+f)}{1+e\cos f}
  \; , \label{di} \\
\end{eqnarray}
where $\omega$ is argument of pericenter, $f$ is the true anomaly at the time of 
impact, and $\eta=\sqrt{1-e^2}$ (e.g., Bertotti et~al. 2003).

Our statistical code does not deal with the detailed geometry of each individual impact. Instead, it follows the 
mean quadratic changes of orbital elements produced by the average effect of impacts with different orientations 
relative to the binary orbit, and times corresponding to different orbital phases. Specifically, we assume that the 
impact direction is isotropic and impacts are not correlated with the binary orbit phase, which should be the case 
for the well-mixed system that we deal with. The algorithm is as follows.

Assuming that ${\bf v}_i = v_i\,{\bf n}$, where $v_i$ is the characteristic impact speed and ${\bf n}$ is the unit 
vector with isotropic orientation, we have 
\begin{equation}
 \left< \left(\delta v_R\right)^2 \right> = \left< \left(\delta v_T\right)^2 \right> =
 \left< \left(\delta v_Z\right)^2 \right> = \frac{1}{3}\,\frac{m'^2}{m^2}\,v_i^2
 \; ,
\end{equation}
and
\begin{equation}
 \left< \delta v_R\,\delta v_T \right> = \left< \delta v_R\,\delta v_Z \right> = 
 \left< \delta v_T\,\delta v_Z \right> = 0 \; ,
\end{equation}
where the average was taken over ${\bf n}$'s orientations. Speed $v_i$ can be determined as the r.m.s. value of 
impact speeds with the distribution that is characteristic to the studied population (e.g., Davis \& Farinella 1997). 

In the next step, we average over the phase of the binary orbit at which the 
impact occurs. We obtain
\begin{equation}
 \frac{\left<\left(\delta a\right)^2\right>}{a^2} = \frac{4}{3}\,\left(\frac{m_i v_i}{m v}
  \right)^2\; , 
\label{a}
\end{equation}
\begin{equation}
 \left<\left(\delta e\right)^2\right> = \frac{5}{6}\,\eta^2\,\left(\frac{m_iv_i}{m v}
  \right)^2\; ,
\label{e} 
\end{equation}
and
\begin{equation}
 \left<\left(\delta \cos I\right)^2\right> = \frac{2+3e^2}{12\eta^2}\,
  \left(\frac{m_iv_i}{m v}\right)^2\; .
\label{i}
\end{equation}
Since the orbital speed of a KBO binary, $v=na$, where $n$ is mean motion, is typically
$\sim$1~m~s$^{-1}$, while $v_i\sim1$ km s$^{-1}$, the binary orbit can be dissolved 
by a single impact generating $\delta a/a$, $\delta e$ or $\delta(\cos I)$ of the 
order of unity, if $m_i/m \gtrsim v/v_i \sim 10^{-3}$. 

In addition, the cumulative effect of impactors with $m_i < 10^{-3} m$ can also be 
important as it leads to a random walk in orbital elements. It can be shown that
\begin{equation}
 \left<\delta a\, \delta e\right> = \left<\delta a\, \delta \left(\cos I\right)
  \right> = \left<\delta e\, \delta \left(\cos I\right)\right> = 0 \; . 
\end{equation}
Thus, the changes in different orbital elements due to small impacts are 
uncorrelated and can be treated separately. 

The binary system is assumed to become dissolved either if $e>1$ or if the semimajor axis 
exceeds some critical value, $a_{\rm crit}$. We set $a_{\rm crit}=0.5 R_{\rm Hill}$
as a rough limit dictated by the Hill stability criterion (e.g., Donnison 2010) and 
numerical integrations (e.g., Nesvorn\'y et al. 2003). By experimenting with the code 
we find that the main channel of binary disruption is reaching $a>a_{\rm crit}$ in one 
or a small number of collisions. Accordingly, the results are insensitive to the
initial distribution of $e$. 

Note that the inclination changes cannot result in binary splitting directly, but, if coupled 
to effects arising from the Kozai dynamics (Kozai 1962), they can also be important. We do not 
consider the coupling of collisional effects with dynamics of inclined binary orbits in this work. 
Therefore, the initial direction of the binary angular momenta does not need to be specified.

\subsection{Boulder with Binary Module}

The binary module was inserted in the {\it Boulder} code. This was done by 
attaching additional data structures to each size bin. These data structures describe
the initial distributions of $a$, $e$ and $i$ of binary orbits, and track how 
these distributions change over time. As the population of impactors evolves with 
time due to collisional fragmentation, the code calculates the time-dependent rate 
of change of binary orbits, and evolves them according to Eqs. (\ref{a})--(\ref{i}). 

\section{Results}

\subsection{PS05 Case}

To start with, we illustrate the case studied by PS05. In this case, the KB
is assumed to evolve in isolation over $\tau=4.5$ Gy. The initial population is given 
a two-slope SFD with $q_<=3$, $q_>=5$, and $R^*$ between 1 km and 50 km. The total 
number of objects with $R>50$~km is normalized to $N_0 = 50,000$, which is roughly 
the estimated number of objects in the present KB. We also consider cases with 
$N = f_N N_0$, where $f_N$ is the scaling parameter. For example, assuming $p_V=0.2$,
$\rho=1$ g cm$^{-3}$, $q_<=3$, $q_>=5$ and $R^*=37$ km, $f_N \sim 0.4$ gives the 
total mass $\approx 0.01 M_{\rm E}$, where $M_{\rm E}$ is the Earth mass, which is what 
the observations seem to indicate for the cold CKB (Fuentes \& Holman 2008), although 
this value is still quite uncertain. On the other hand, $f_N = 13$ corresponds to 
the population used in PS05 (see discussion in \S1).

\begin{figure}
\epsscale{0.6}
\plotone{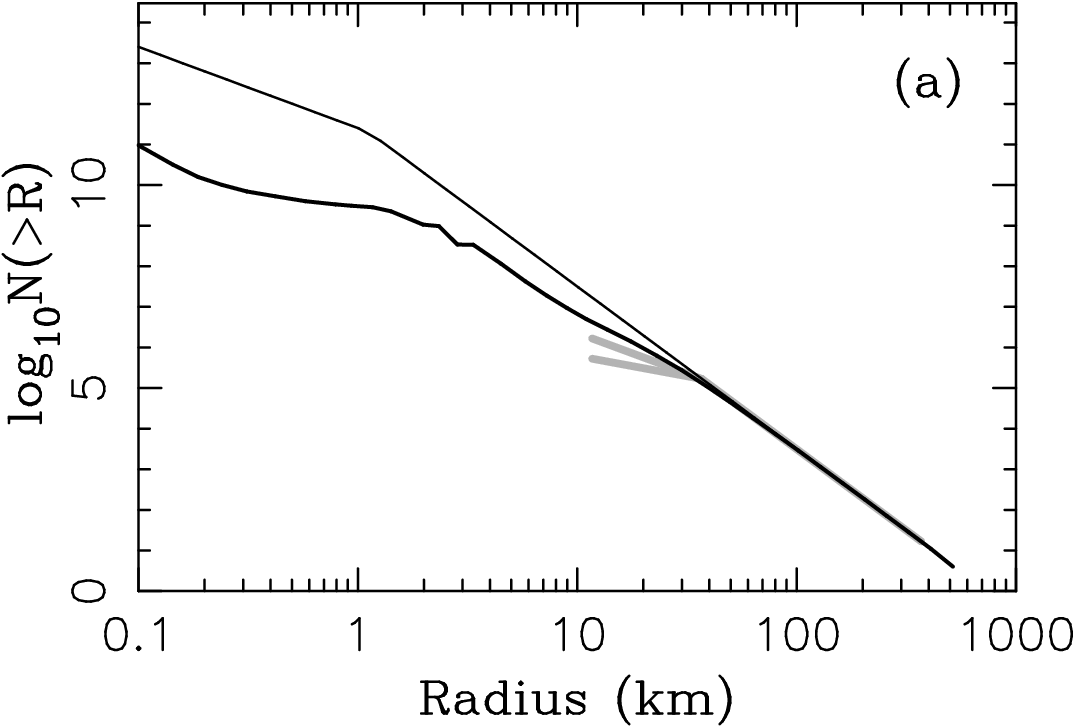}\vspace*{5.mm}
\plotone{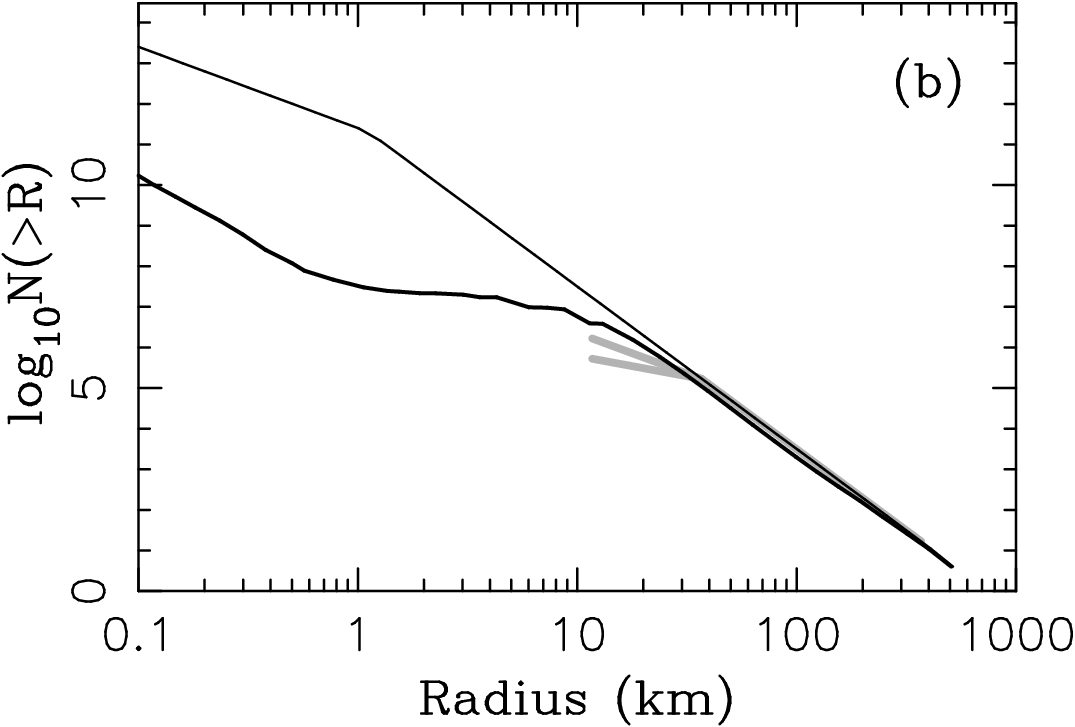}
\caption{Kuiper belt SFD changes for: (a) $f_Q=3$, and (b) $f_Q=500$. These two fragmentation laws 
roughly correspond to the strong and weak disruption laws considered by PS05 ($\beta=3$ and $\beta=3/2$,
respectively, in their notation). The upper thin black lines in both panels show the initial
SFD ($f_N=1$). The bottom bold black lines are the SFD produced by collisional grinding over 
$\tau = 4.5$ Gy. The bold grey lines denote constraints on the present SFD of the cold CKB: 
$q_>=5$ for $R>37$ km according to Fraser et al. (2010) and $q_<=2$-3 for $R<37$~km for according 
to Fuentes et al. (2008, 2009). Since we plot the cumulative distributions $N(>R)$ here, the 
plotted slopes have indices equal to 4 for $R>37$~km and 1-2 for $R<37$~km.}
\label{ps05}
\end{figure}

Figure \ref{ps05} shows the result of collisional grinding for $f_N=1$ and two different 
values of $f_Q$: $f_Q=3$, roughly corresponding to the PC05's strong disruption law, and 
$f_Q=500$, roughly corresponding to the PC05's weak disruption law (see Fig. \ref{qstar}). 
We used $P_i=4 \times 10^{-22}$~km$^{-2}$~yr$^{-1}$ and $v_i=1$ km s$^{-1}$.
The initial distribution was set so that $q_<=3$ for $R<1$ km and $q_>=5$ for $R>1$ km.
For $\rho=1$ g cm$^{-3}$ this gives the total initial mass of 0.84 $M_{\rm E}$. After having
evolved the population with the {\it Boulder} code over $\tau=4.5$ Gy, the remaining 
masses were 0.056 $M_{\rm E}$ for $f_Q=3$ (Fig. \ref{ps05}a) and 0.027 $M_{\rm E}$ for 
$f_Q=500$ (Fig. \ref{ps05}b). 

For $f_Q=3$, the SFD slope just below $\sim$50 km becomes slightly shallower over time, 
mainly due to the effects of large cratering impacts, and reaches $q\sim4$ at $\tau=4.5$ Gy. 
This final slope is significantly steeper than the present slope of cold CKBOs at these radii. 
With $f_Q=500$, a sharp roll-over to a very shallow slope develops at $R\sim10$ km, because
most objects with $R<10$ km suffer catastrophic disruptions. Thus, even with the unrealistically 
weak disruption law, the SFD break is still significantly below the actual roll-over 
radius in the present cold CKB. This result is insensitive to the specific choice of model 
parameters related to the generation of fragments, resolution, and plausible changes of 
$P_i$ and/or $v_i$. 

We therefore find that the observed break at $R^*=25$-50 km cannot be produced by collisional 
grinding, except if the number of objects with $R>R^*$ was much larger in the past, and 
was depleted by dynamical processes, or if the cold CKBO overlapped with a much larger 
population of impactors in the past. We consider these possibilities in the following 
text. 
 
\subsection{Adopted Model}

Levison et al. (2008a) proposed that most of the complex orbital structure seen in 
the KB region today (see, e.g., Gladman et al. 2008) can be explained if bodies native 
to 15-35~AU were scattered to $>$35 AU by eccentric Neptune (Tsiganis et al. 2005). 
If these outer solar system events coincided in time with the Late Heavy Bombardment (LHB) 
in the inner solar system (Gomes et al. 2005), binaries populating the original planetesimal 
disk at 15-35~AU would have to withstand $\sim$600 My before being scattered into the Kuiper 
belt. Even though their survival during this epoch is difficult to evaluate, due to major 
uncertainties in the disk's mass, SFD and radial profile, the near-absence of wide and equal-sized 
binaries among 100-km-sized {\it hot} classical KBOs (Noll et al. 2008a,b) seems to indicate 
that the unbinding collisions and scattering events must have been rather damaging. 

The contrasting characteristics of {\it cold} CKBOs discussed in \S1 may indicate 
that the cold CKBOs formed in a relatively quiescent environment at $>$40 AU rather than having 
been scattered to their current orbits from $<$35~AU, 
because more resemblance between different trans-Neptunian populations 
would be expected in the latter model. The in-situ formation of cold CKBOs is also 
supported by the results of Parker \& Kavelaars (2010) who showed that some of the widest binaries 
observed in the cold CKB today would probably not survive scattering encounters with Neptune.  

To understand the collisional history of cold CKB, we consider both the pre-LHB and post-LHB 
epochs. For the pre-LHB epoch, we assume that the cold CKBOs evolved in relative 
isolation at 40-50 AU, where most collisions occurred between cold CKBOs themselves. The 
results of these pre-LHB simulations should also apply, with minor modifications, if the cold 
CKBOs formed closer in, as far as the source population of cold CKBOs can be considered as a 
closed collisional system. The coupling of CKBOs to the scattered trans-planetary disk 
{\it during} the LHB is not considered here because Benavidez \& Campo Bagatin (2009) showed that 
collisional fragmentation during this stage was unlikely to produce substantial changes 
in the SFD.

\subsection{Post-LHB Epoch}

We start by discussing the collisional evolution of the Kuiper belt after LHB. We consider 
collisions between two CKBOs, one CKBO and one scattered disk object, and two scattered 
disk objects. The scattered disk is massive initially and becomes dynamically depleted over 
time, with an estimated depletion factor of 100-250 over 4 Gy (Levison \& Duncan 1997, 
Dones et al. 2004, Tsiganis et al. 2005). This dynamical depletion is taken into account in 
{\it Boulder} by gradually decreasing the number of scattered disk objects in all size bins.

The collisional probabilities, impact speeds, initial SFDs and dynamical decay rates 
were taken from Levison et al. (2008b). 
Specifically, we used $P_i=4\times10^{-22}$ km$^{-2}$ yr$^{-1}$ and $v_i=1$ km s$^{-1}$ for 
CKB collisions, $P_i=2\times10^{-22}$ km$^{-2}$ yr$^{-1}$ and $v_i=1.5$ km s$^{-1}$ for 
CKB--scattered-disk collisions, and $P_i=10^{-22}$ km$^{-2}$ yr$^{-1}$ and $v_i=3$ km s$^{-1}$ 
for collisions between scattered disk objects (see also Brown et al. 2007). We 
varied these (and other) parameters in test simulations to determine the sensitivity of 
results to different assumptions. 

The SFD of the scattered disk was normalized as in Levison et al. (2008b). We assumed 
that there are presently 50,000 scattered disk objects with $R>50$ km, and that this 
population decayed by a factor of 250 times since LHB (Tsiganis et al. 2005). We fixed 
$f_N=1$, because the dynamical depletion of CKBOs should be relatively minor during this 
stage. For $R^*=37$ km, this gives the initial mass $\sim$0.2 $M_{\rm E}$. Also, to promote 
collisional grinding, we used the weak disruption law with $f_Q=10$. 

\begin{figure}
\epsscale{0.6}
\plotone{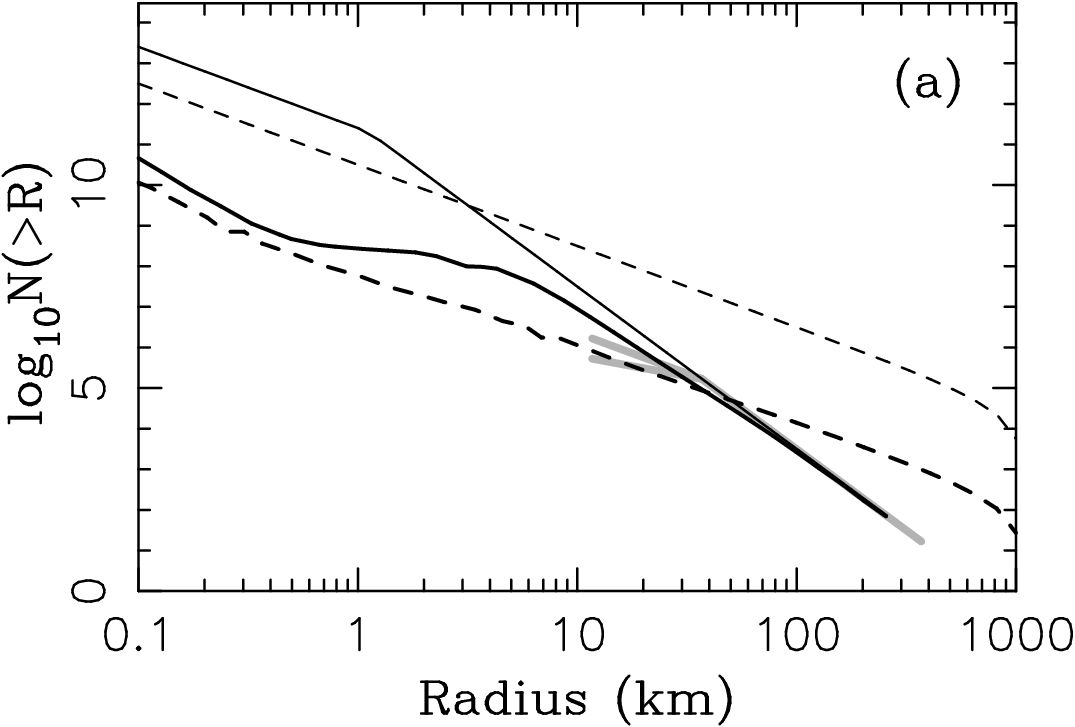}\vspace*{5.mm}
\plotone{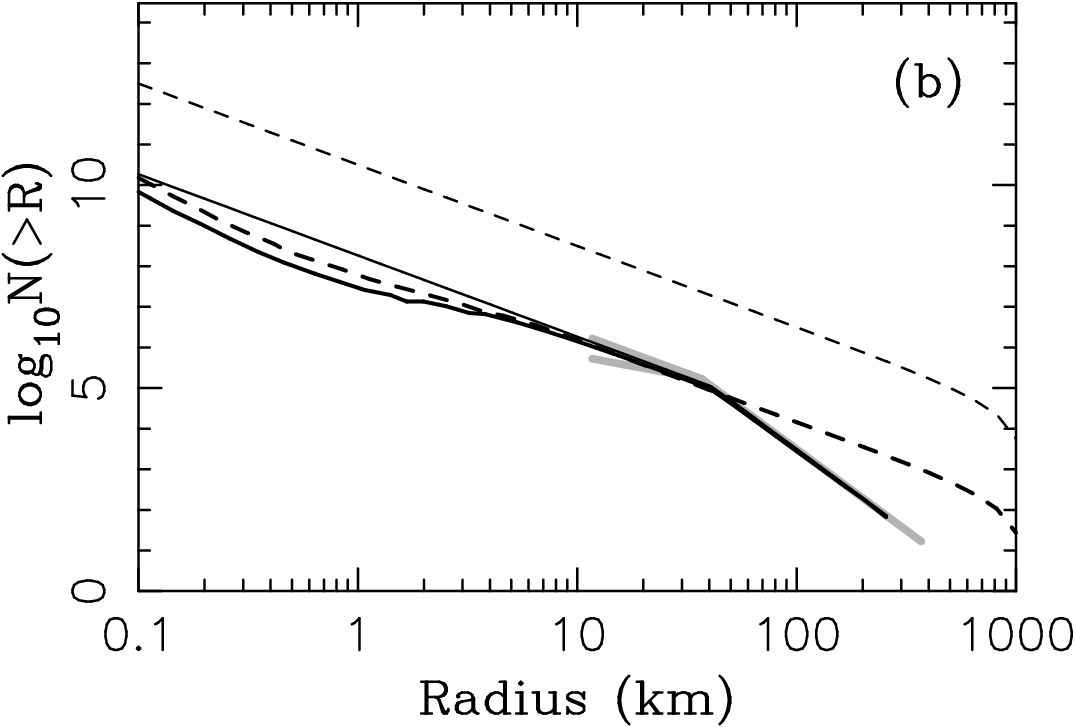}
\caption{SFDs of CKBOs (solid lines) and scattered disk objects (dashed lines) in our {\it Boulder}
simulations of the post-LHB epoch for: (a) the initial SFD of CKBOs steep for $R>1$ km ($q=5$) 
and shallow for $R<1$ km ($q=3$), and (b) initial SFD of CKBOs with a break at $R^*=37$ km. In both 
cases, we assumed that the initially massive scattered disk dynamically decayed as in  
Tsiganis et al. (2005). The thin lines show the initial distributions. The bold lines 
are the final SFDs evolved over $\tau = 4$ Gy. We used $f_Q=10$ and $f_N=1$.}
\label{scat}
\end{figure}

Figure \ref{scat} shows the SFD of CKBOs and scattered disk objects that were obtained for
two different assumptions on the initial SFD of cold CKB. In Fig. \ref{scat}a, we assumed 
that the initial SFD was steep down to $R=1$ km. In Fig. \ref{scat}b, we used the present SFD 
shape of cold CKB with $R^*=37$ km. 

In Fig. \ref{scat}a, the roll-over in the final SFD of CKBOs occurs at $R\approx5$ km,
which is significantly below the observed $R^*$ value. We experimented with a range of $f_N$, 
$f_Q$, $v_i$, and $P_i$ values, and various initial SFDs. These tests showed that the results
illustrated in Fig. \ref{scat}a are representative. Specifically, if we start with the
initial break at $R<10$ km, the final break ends up being at $R<10$ km as well. These results 
therefore suggest that the observed SFD roll-over of cold CKBOs at $R^*=25$-50 km cannot 
be produced by the collisional grinding {\it after} LHB. 
  
On the other hand, if we start with the break at $R^*=37$ km (Fig. \ref{scat}b), the SFD of 
CKBOs remains nearly unchanged for $R\gtrsim10$ km. The total mass loss due 
to the collisional grinding of small objects is only $\sim$15\% in the CKB and $\sim$5\% in 
the scattered disk. These results show that, while the SFD may have been shaped by collisional 
grinding in epochs prior to the LHB, it remained essentially constant since then. The 
binary fraction does not provide any interesting constraints on the collisional evolution 
of CKBOs after LHB because the vast majority of binaries with $R\gtrsim10$ km survive. 

\subsection{Pre-LHB Epoch}

Fraser (2009) and Benavidez \& Campo Bagatin (2009) suggested that the observed SFD roll-over in 
the cold CKB was produced by collisional grinding during the $\sim$600 My before the LHB. The collisional
modeling of the pre-LHB phase is complicated by the fact that the state of the Kuiper belt before 
LHB is poorly understood. What we seem to infer from observations (see discussion in \S1 and \S3.2) 
is that the cold CKB probably formed in situ at 40-50 AU. The LHB modeling would then 
indicate that various dynamical processes probably removed $\sim$90\% of its mass during 
LHB (Morbidelli et al. 2008). Using these results as a guideline, we find that there is indeed a 
potential for the observed SFD roll-over being a fossil remnant of the collisional grinding of 
massive population before LHB.

The impact speeds in the pre-LHB CKB depend on the dynamical state of the disk. In its 
present state, the impact speeds are relatively high ($\sim$1 km s$^{-1}$) because orbits 
have relatively high eccentricities ($e\sim0$-0.2) and inclinations ($i\sim0$-30$^\circ$).   
On the other hand, KBOs can only form if $e$ and $i$ were much lower (e.g., Kenyon et al. 2008 
and the references therein). This implies that some dynamical processes must have excited
$e$ and $i$ to their present values. Two main possibilities exist. Either (1) the cold CKB was 
dynamically excited by some primordial process that dates back to KBO formation, or (2) the 
excitation was produced by the LHB itself (see Morbidelli et al. 2008 for a discussion). Following 
Fraser (2009) we first study (1), in which case $v_i\sim 1$ km s$^{-1}$. Low $v_i$ values implied 
by (2) will be discussed in \S3.7.

Fraser (2009) considered the initial SFDs that were steep ($q_1=5$) down to $R_a$, then became 
nearly flat ($q_2=0$-2) down to $R_b$, with the collisional equilibrium below $R_b$ ($q_3=3.5$). 
These initial SFDs were motivated by the results of published simulations of the coagulation 
growth in KB, which tend to produce such distributions (Kenyon et al. 2008). Since it is 
not clear, however, whether KBOs formed by two-body coagulation in the first place (see Chiang 
\& Youdin 2010 for a review), it is not guaranteed that these initial conditions actually 
apply. Nevertheless, we use Fraser's initial SFDs as a starting point and consider other 
options in \S3.6.

\begin{figure}
\epsscale{0.6}
\plotone{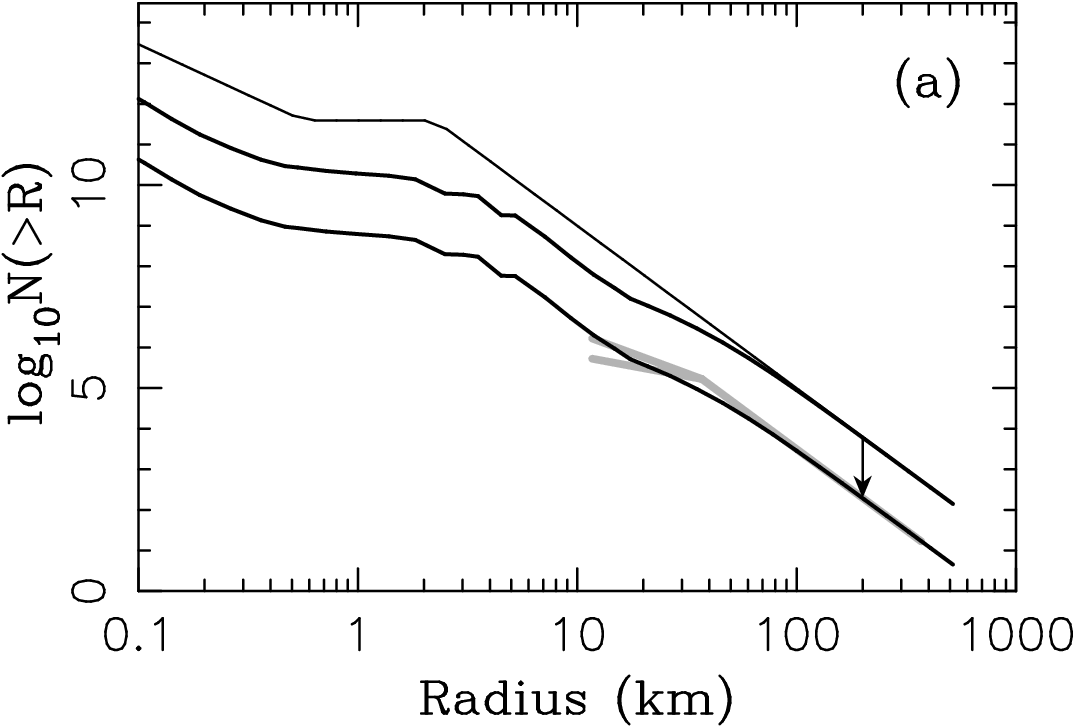}\vspace*{5.mm}
\plotone{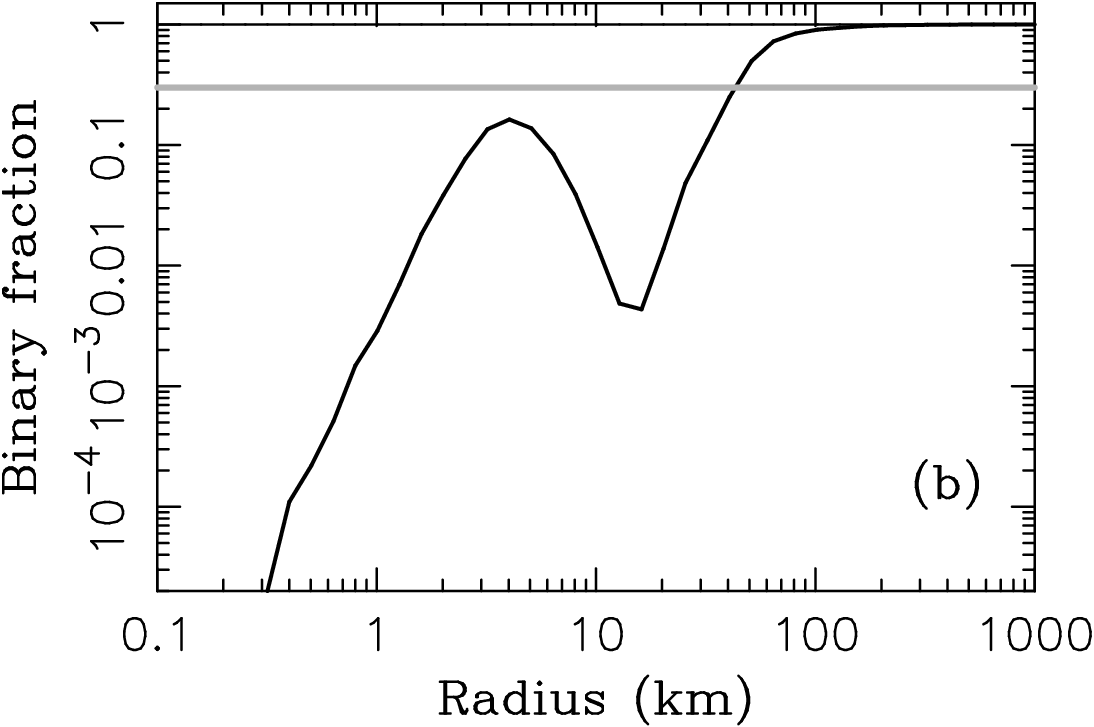}
\caption{(a) SFD evolution and (b) binary survival before LHB for the initial SFD shape used by 
Fraser (2009): $R_a=2$ km, $R_b=0.5$ km, $q_1=5$, $q_2=1$ and $q_3=3.5$. We used 
$f_Q=3$ corresponding to the Fraser's weak disruption law. The thin black line in (a) shows the 
initial SFD. The middle bold black line in (a) shows the final state after $\tau = 600$ Myr of
collisional grinding. The bottom bold black line in (a) shows the SFD after a dynamical depletion 
factor $f_N=30$ was applied to the final distribution (change indicated by an arrow). The bold grey 
lines in (a) denote observational constraints on the present SFD in cold CKB. The bold grey line 
in (b) marks the binary fraction of 0.3. We assumed that the initial binary fraction was 1. The bold 
black line in (b) shows the expected binary fraction in this model. The binary separation was 
set to 0.01 $R_{\rm H}$.}
\label{fraser}
\end{figure}

Figure \ref{fraser} illustrates the Fraser's case with $R_a=2$ km, $R_b=0.5$ km, $q_1=5$, $q_2=1$
and $q_3=3.5$. As in Fraser (2009), we used $v_i=1$ km s$^{-1}$, $f_Q=3$ (corresponding 
to Fraser's weak disruption law), and evolved the populations with {\it Boulder} over $\tau=600$ Myr. 
Factor $f_N$ was varied to obtain different initial masses and thus different collisional 
histories. Each of these cases would imply a different dynamical depletion factor during LHB.

The best results were obtained with $f_N\sim$20-50, implying the initial mass between 7 and 15 
$M_{\rm E}$ (Fig. \ref{fraser}a). With $f_N<20$ ($<$7 $M_{\rm E}$) and $f_N>50$ ($>$15 $M_{\rm E}$), 
the SFD roll-over occurred at radii that were either too small and too large, respectively, 
compared to observations. These values are only soft limits, however, because the roll-over radius 
is also sensitive to the assumed $R_a$ value. For example, smaller initial mass values would still 
be plausible if $R_a=3$-10 km. In addition, slightly larger roll-over radii can be produced 
with $f_Q=10$. We therefore find, in agreement with Fraser (2009), that a reasonably conservative 
lower limit on the initial disk's mass is $\sim$1 $M_{\rm E}$. 

The model SFDs discussed here share a common trait. While they bend to a shallow slope at $R^*$, 
the slope below $R^*$ is never shallower than $q\approx3$ and, even in the best cases such 
as the one shown in Fig. \ref{fraser}a, just barely matches the observational constraint. Moreover,
the shallow SFD segment below $R^*$ generally only extends down to $R=10$-20 km, and steepens 
back to $q\approx5$ for $R\lesssim10$ km (Fig. \ref{fraser}a). This does not contradict the existing 
observations because very little is known about KBOs with $R\lesssim10$ km. We were unable obtain 
a case where the final SFD would be uniformly shallow below $R^*$ down to $R<10$ km, except for 
very large and probably implausible initial masses. This is therefore clearly not the idealized 
case considered in PS05, where it was assumed that the slope below $R^*$ can be approximated by 
$q\approx3$ to some very small (indefinite) $R$. 

Our simulations show that the disk mass is reduced by a factor of $\sim$10 by collisional grinding.
Thus, starting with $\sim$1 $M_{\rm E}$, the remnant mass just before LHB would be $\sim$0.1~$M_{\rm E}$.
This is plausible because the LHB modeling in Morbidelli et al. (2008) showed that the population at 
40-50 AU becomes dynamically depleted by a factor of $\sim$10. The expected final mass of cold CKBOs 
from these order of magnitude considerations is therefore $\sim$0.01~$M_{\rm E}$, which is in the right 
ballpark when compared to observations (e.g., Fuentes \& Holman 2008, Brucker et al. 2009). 
 
The initial masses smaller than $\sim$1 $M_{\rm E}$ would imply $R_a>10$ km, and indicate that the 
observed break at $R^*\sim37$ km was essentially in place before fragmentation processes had begun.
The initial masses larger than $\sim$10 $M_{\rm E}$ are probably implausible, because these large masses 
would imply $\gtrsim$1 $M_{\rm E}$ mass before LHB and would require a dynamical depletion 
factor at LHB in excess of $\sim$100. Such a large depletion factor is difficult to explain by LHB
processes if the cold CKBOs formed in situ at 40-50 AU (Morbidelli et al. 2008). 

\subsection{Binary Survival}

We now consider the binary survival. Figure \ref{fraser}b shows the fraction of binaries 
surviving the pre-LHB epoch for $f_N=30$ and Fraser's initial SFD shape. There is a clear trend 
with the physical size of binary components. Specifically, more than 50\% of binaries with radii 
$R>50$ km survive, while the survival rate for $R=10$-20 km is only $\sim$0.5\%. This trend is 
easy to understand because, according to Eqs. (\ref{a})--(\ref{i}), smaller binary mass $m$ 
implies larger orbital change. 

The trend is reversed for $R<10$ km because of the lack of small impactors with $R\sim0.5$~km 
that could unbind binaries with $R\sim5$ km (see Fig. \ref{fraser}a). The behavior of the 
surviving binary fraction for $R<10$ km is sensitive to the initial SFD. For example, if 
$R_a=R_b=2$~km, in which case the Dohnanyi's slope is directly attached to the steep SFD slope at 
large sizes, the survival rate of $R<10$ km binaries more monotonically drops with decreasing $R$
(see \S3.6). We will concentrate on binaries with $R>10$ km in the following discussion, because 
that's where things can be constrained by the existing observational data.

\begin{figure}
\epsscale{0.6}
\plotone{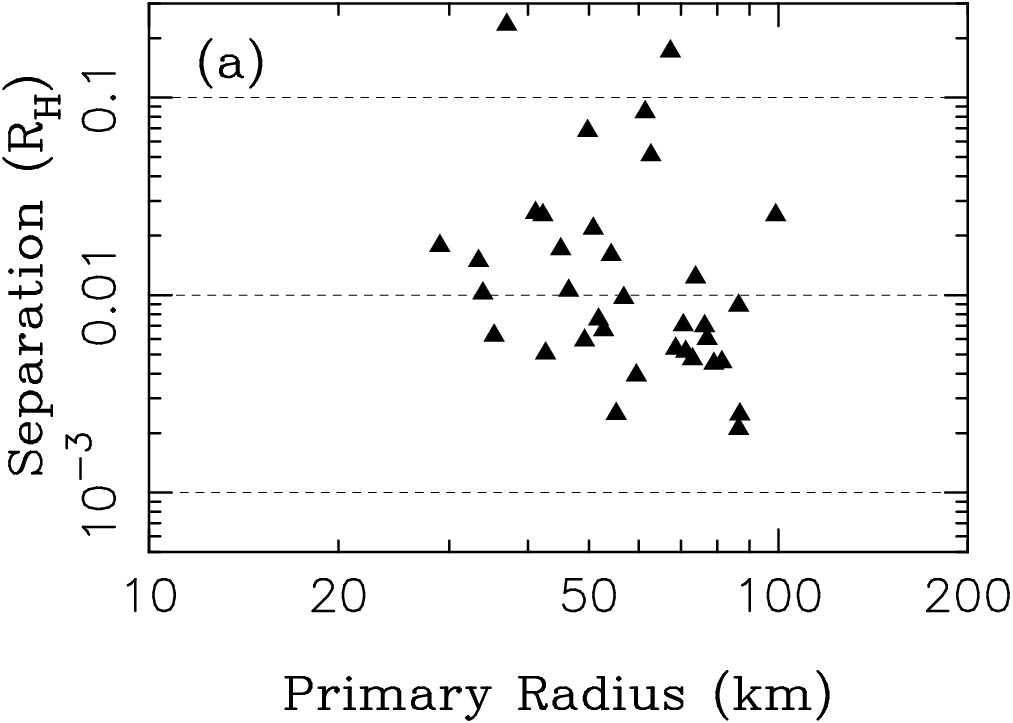}\vspace*{5.mm}
\plotone{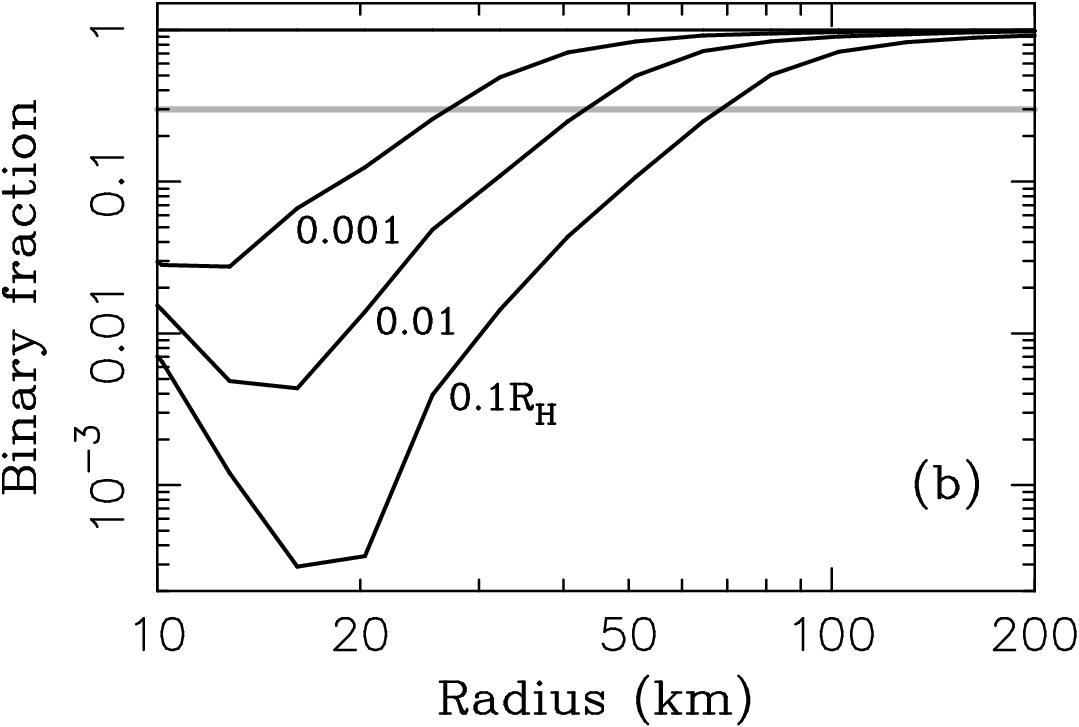}
\caption{(a) Physical parameters of known binaries in the cold CKB (Noll et al. 2008a). Separations 
were computed from the apparent angular separation of the two components at discovery. Radius of 
the primary was estimated with $p_V=0.2$. The horizontal dashed lines in (a) denote the separation 
values considered in our model. (b) The same as Fig. \ref{fraser}b, but for three different initial 
separations of binary components. As expected, the wide binaries have lower survival rates than the 
tight ones.} 
\label{corel}
\end{figure}

Figure \ref{corel}a shows the physical parameters of known binaries in the cold CKB (Noll et al. 
2008a, Grundy et al. 2009). The radii of primary components range from $\sim$30 to $\sim$100 
km. The separations are between $\sim$$2\times10^{-3}$ to $\sim$0.2 Hill sphere. From Eq. (\ref{dv}), 
the survival rate should decrease with separation because $\delta v/v \propto v_i \sqrt{a}$, which
is larger for larger $a$. Given the spread of separations in Fig. \ref{corel}a, we therefore  
consider cases with $a=0.001$, 0.01 and 0.1 $R_{\rm H}$. These initial values should cover the 
interesting range of separations. Considering these cases separately, rather than using some 
continuous initial distribution of $a$, is the right thing to do, because the real distribution 
of $a$ produced by the formation process is not well understood.

Figure \ref{corel}b shows the surviving binary fraction for these semimajor axis values. As expected,
the wider binaries have lower survival rates than the tighter 
ones. For example, $R=30$ km binaries with $a=0.1$ $R_{\rm H}$ would be reduced by a factor of 
$\sim$100 in the Fraser's pre-LHB collisional model illustrated in Fig. \ref{fraser}a,
while $R=30$~km binaries with $a=0.01$ $R_{\rm H}$ would be reduced by a factor of $\sim$10. 
These considerations provide an interesting test on the level of collisional grinding in the
cold CKB, because the binary fraction could have been strongly reduced by collisions, especially 
for $R\lesssim50$ km. To pass this test, any plausible collisional scenario needs to match not only 
the SFD of cold CKBOs, including the roll-over at $R^*=25$-50 km, but also explain how the large 
fraction of cold CKBOs binaries survived, as indicated by observations (see \S4). 

\subsection{Sensitivity to Initial SFD}

\begin{figure}
\epsscale{0.6}
\plotone{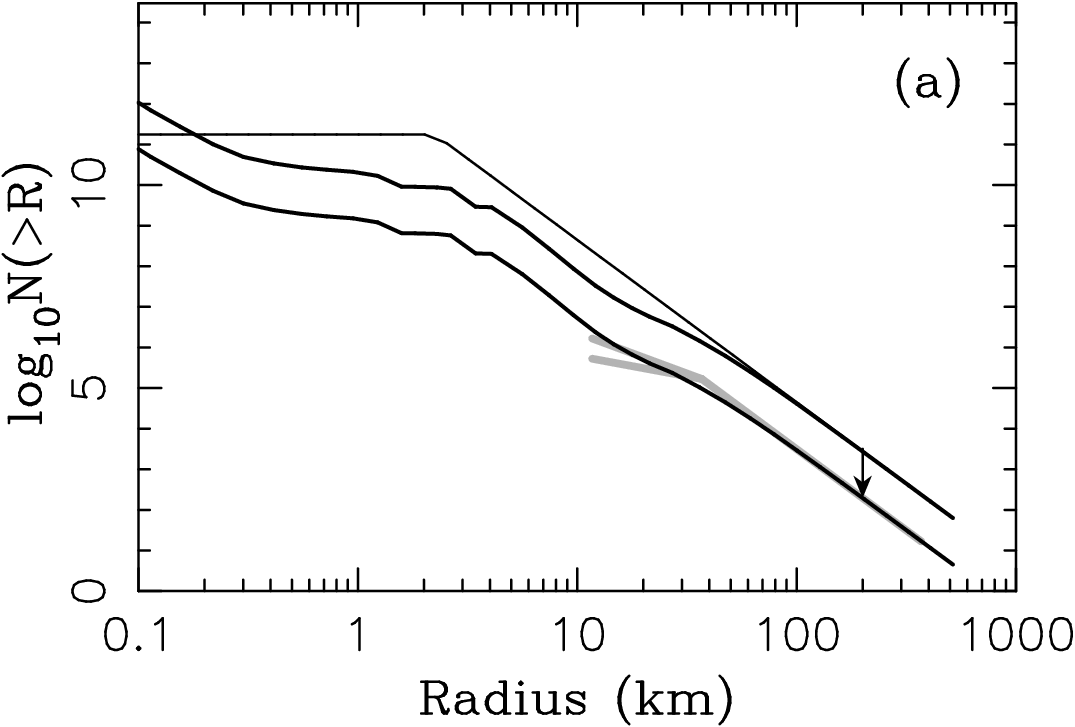}\vspace*{5.mm}
\plotone{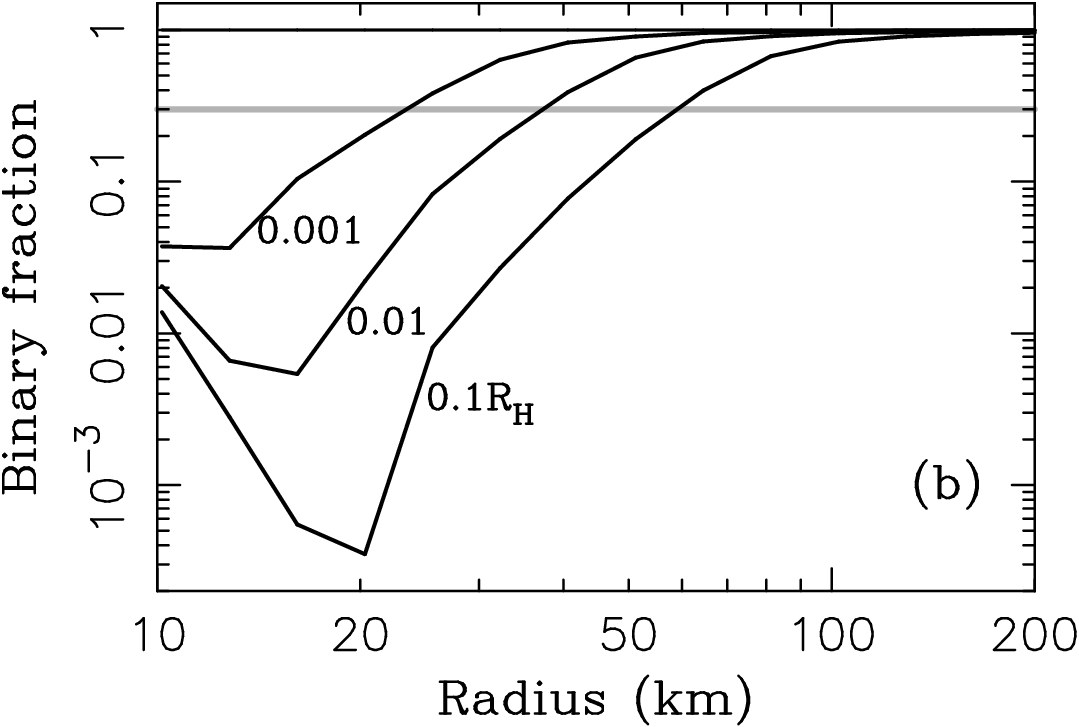}
\caption{The same as Fig. \ref{fraser} but for the initial SFD that lacks objects with $R<2$ km. 
We set $f_N=14$ corresponding to the initial mass of $\approx$4 $M_{\rm E}$. The initial population 
grinds down to $\approx$0.6 $M_{\rm E}$ in $\tau=600$ Myr before LHB.}
\label{endm1}
\end{figure}

\begin{figure}
\epsscale{0.6}
\plotone{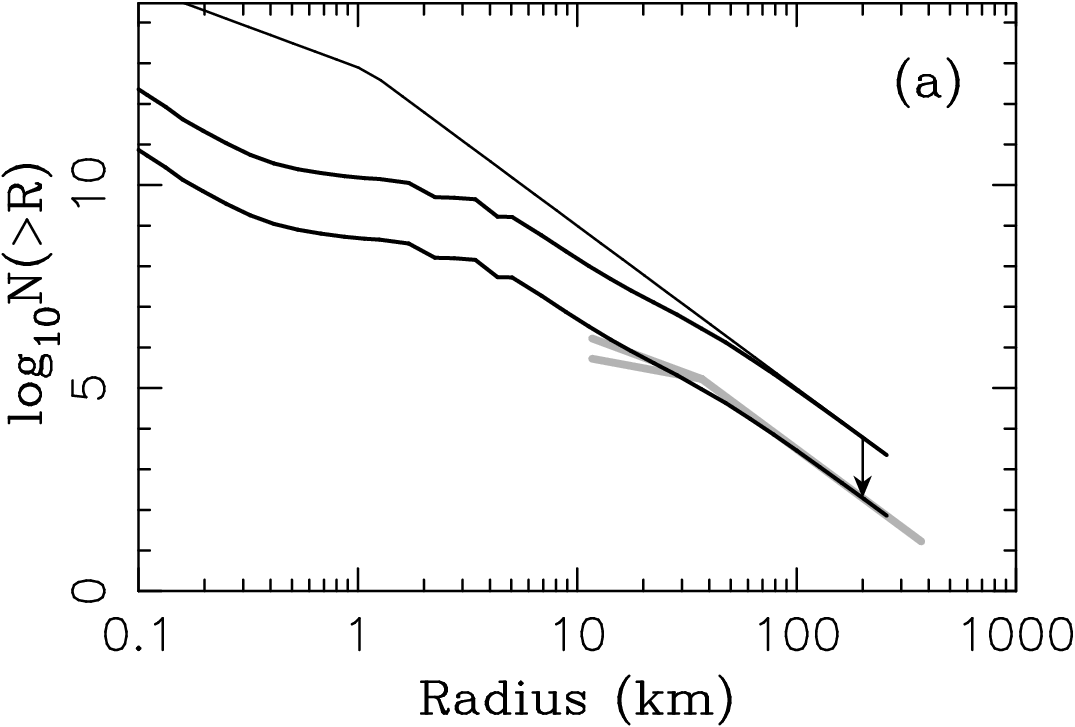}\vspace*{5.mm}
\plotone{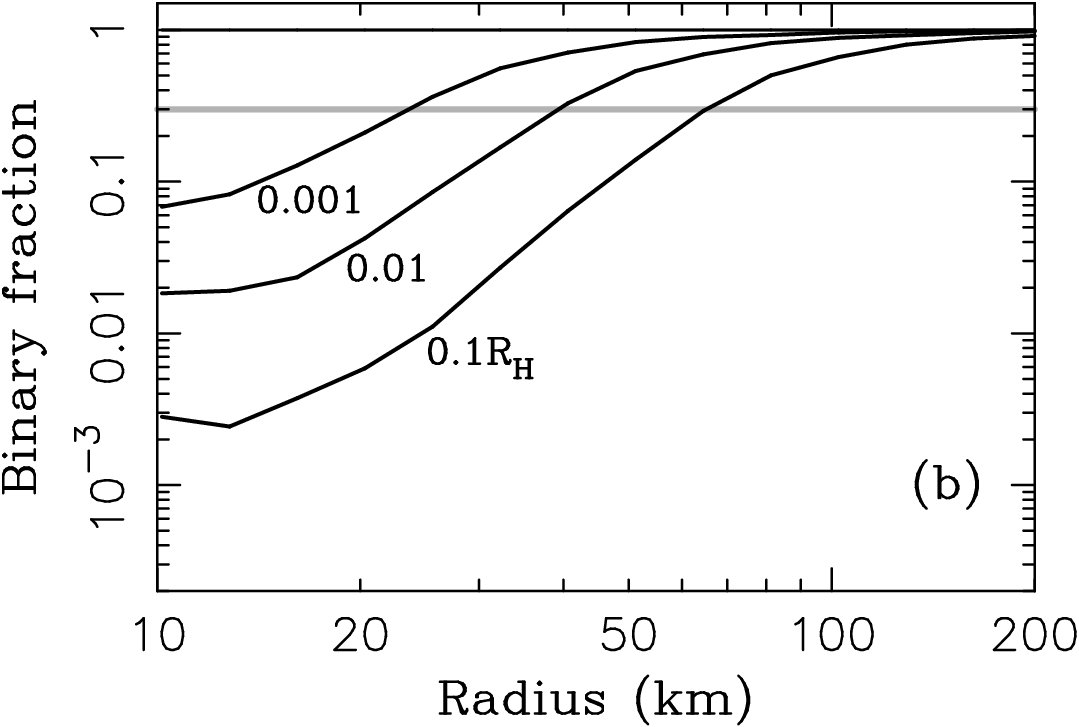}
\caption{The same as Fig. \ref{fraser} but for the initial SFD with a break from $q=5$ to $q=3$ at 
$R=2$ km. We set $f_N=30$ corresponding to the initial mass of $\approx$26 $M_{\rm E}$. The initial 
population grinds down to $\approx$1 $M_{\rm E}$ in $\tau=600$ Myr before LHB.}
\label{endm2}
\end{figure}

Before we compare our results to observational data, we test the sensitivity of the binary survival 
to different assumptions. Figures \ref{endm1} and \ref{endm2} illustrate
the dependence on the initial SFD. In Fig. \ref{endm1}, we choose to extend the steep slope
($q=5$) from large sizes down to $R=2$~km, and assume that there are no bodies initially with 
$R<2$ km (case A). In Fig. \ref{endm2}, we impose a break from $q=5$ to $q=3$ at $R=2$ km (case B). 
In each case, $f_N$, or equivalently the total initial mass, is set so that the collisional grinding
leads to the SFD roll-over at $R=25$-50 km. The population is evolved over $\tau=600$ My.

The best results were obtained with $f_N=14$ (4 $M_{\rm E}$ initial mass) for case A and $f_N=30$ 
(26 $M_{\rm E}$) for case B. These initial masses grind down to 0.6 and 1 $M_{\rm E}$, respectively.   
While the final distribution obtained in case A matches constraints reasonably well (Fig. \ref{endm1}a,
but see discussion in \S3.4), the distribution for $R<30$ km in case B is always steeper than $q=3$. 
Interestingly, it is difficult to produce a shallower slope in case B unless we use the initial 
masses in excess of 100 $M_{\rm E}$, which is clearly implausible, or $f_Q \gg 10$, which would 
conflict with the published results of impact simulations (see \S3.1).  

It thus appears that a rather abrupt change in the {\it initial} SFD slope is needed to produce a 
shallow slope with $q=2$-3 below $R^*$. This can be easily understood. The objects with $R$ near the 
initial slope change are long-lived, because they see a small number of impactors, if the transition
is sharp. Thus, they can break larger bodies and create a sharp SFD roll-over at about ten 
times their radius. We find that it is easier to fit constraints if the initial break is placed at 
$R\sim5$ km, rather than at $R\lesssim2$ km, because $R\sim5$ km objects are long-lived in that 
case and can disrupt KBOs near $R^*$ with our disruption laws ($f_Q=1$-10).  

Additional dependencies exist on the assumed SFDs of fragments produced by the cratering
and catastrophic impacts. We find that catastrophic impacts tend to produce a sharp transition
from steep to shallow slope near the largest object in the population that can be disrupted by 
them. The cratering impacts, on the other hand, tend to smooth this transition and produce more
gentle waves in the SFD. This may explain some of the subtle differences between our simulations, 
which tend to produce gentle SFD waves, and those of Fraser (2009), which show stronger 
variations of slope for $R \sim R^*$ (e.g., $q<2$ below the break).   

The fraction of binaries surviving in case A is within a factor of $\sim$2 to that we obtained with 
the Fraser's initial distribution. Interestingly, the binary survival rate is slightly larger in case B, 
which has larger mass than case A, and should thus lead to more perturbations on binary orbits. The
larger binary survival rate in case B is related to the shorter lifetime of intermediate-size
impactors ($R\sim5$ km) in case B, which are disrupted by smaller impactors from the Dohnanyi's tail.
These small impactors are nearly absent in case A.  

\subsection{Sensitivity to Impact Speeds}

So far we described tests with $v_i\sim1$ km s$^{-1}$. It is uncertain whether these assumed impact 
speeds should apply to the pre-LHB collisions, because the dynamical state of the Kuiper belt region 
at 40-50 AU could have been very different. Specifically, the dynamical effects during LHB such as, 
e.g., passing resonances, could have excited the orbits of cold CKBOs. If so, the cold CKBOs could 
have had smaller eccentricities and inclinations than they have now, implying lower collision speeds before the 
LHB. Here we test cases with $v_i<1$~km~s$^{-1}$. The results need to be considered with caution, because 
it is not clear whether our disruption laws (\S2.1) are applicable with low impact speeds.

\begin{figure}
\epsscale{0.6}
\plotone{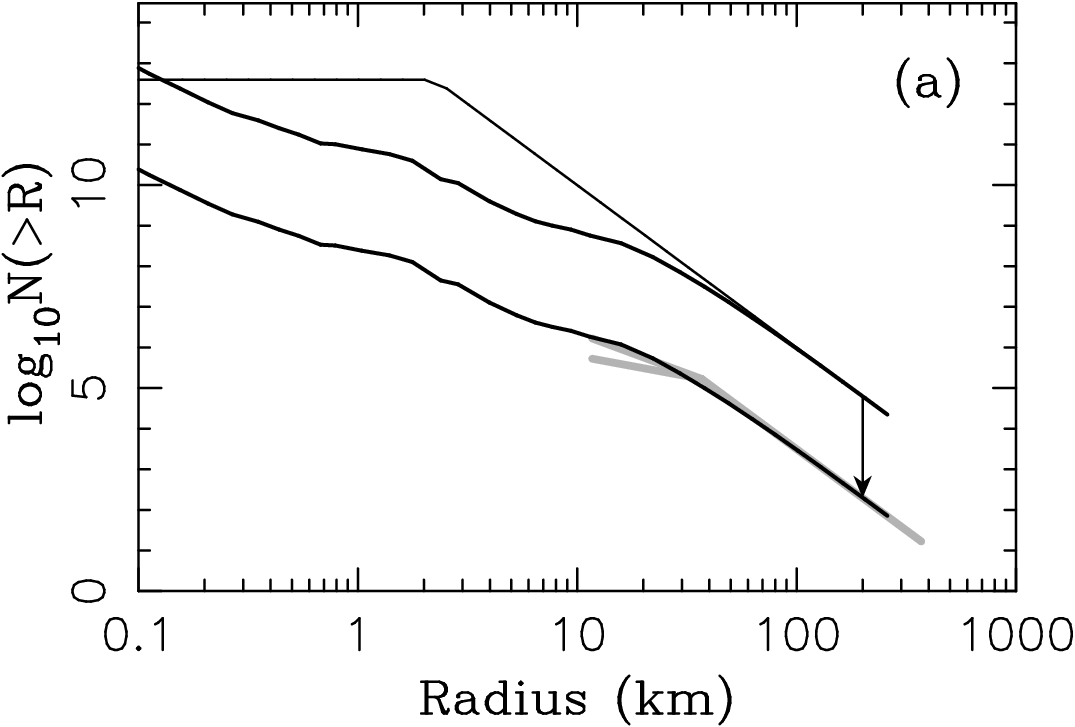}\vspace*{5.mm}
\plotone{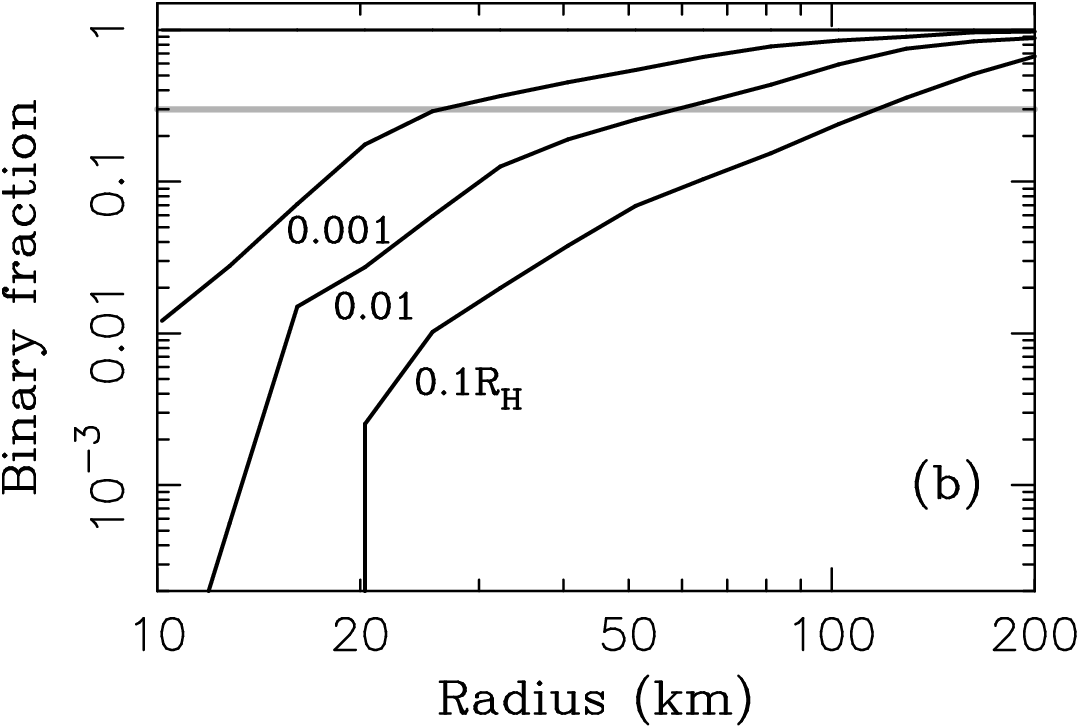}
\caption{The same as Fig. \ref{endm1} but for $v_i=300$ m s$^{-1}$. We set $f_N=300$ corresponding to the 
initial mass of $\approx$88 $M_{\rm E}$. The initial population grinds down to $\approx$6.7 $M_{\rm E}$ in 
$\tau=600$ Myr before LHB.}
\label{low1}
\end{figure}

Figure \ref{low1} illustrates the case with $v_i=300$ m s$^{-1}$. To create the roll-over at $R^*$
with this low collisional speed, we needed to assume $f_N=300$, giving the initial mass of 88 $M_{\rm E}$ 
for the case-A initial SFD. The population grinds down to 6.7 $M_{\rm E}$ at $\tau=600$ My, which
would imply a dynamical depletion factor of $>$200. Collision speeds $v_i<300$ m s$^{-1}$ would require 
even larger initial masses and dynamical depletion factors that are clearly implausible. We therefore 
find that it is difficult to produce the SFD roll-over by collisional grinding with low collisional 
speeds ($v_i\lesssim300$ m s$^{-1}$). 

The surviving binary fraction for $v_i=300$ m s$^{-1}$ (Fig. \ref{low1}b) shows the usual trend with 
$R$ in that the binaries with physically smaller components survive at lower rate than the larger ones. 
Below $R=10$-20 km, the surviving binary fraction drops to $<$$10^{-3}$. The results for $R>20$ km are 
similar to those obtained with $v_i\sim1$ km s$^{-1}$ indicating that the survival of larger binaries 
is not overly sensitive to the assumption on $v_i$.  
 
Additional tests show that it might be more plausible to create the roll-over at $R^*$ with low $v_i$, 
if $r_1\sim5$ km (initial 20 $M_{\rm E}$ grinds to 5 $M_{\rm E}$) and/or 
for $f_Q=10$ (20 $M_{\rm E}$ grinds to 2 $M_{\rm E}$). The results for binary survival 
are similar in these two cases. While $r_1=5$ km generates a slightly stronger gradient with $R$ and 
minimum survivability for $R<20$ km, $f_Q=10$ produces a slightly softer gradient that is 
similar to that in Fig. \ref{endm1}b. 

\subsection{Changes of Binary Orbits} 

\begin{figure}
\epsscale{0.6}
\plotone{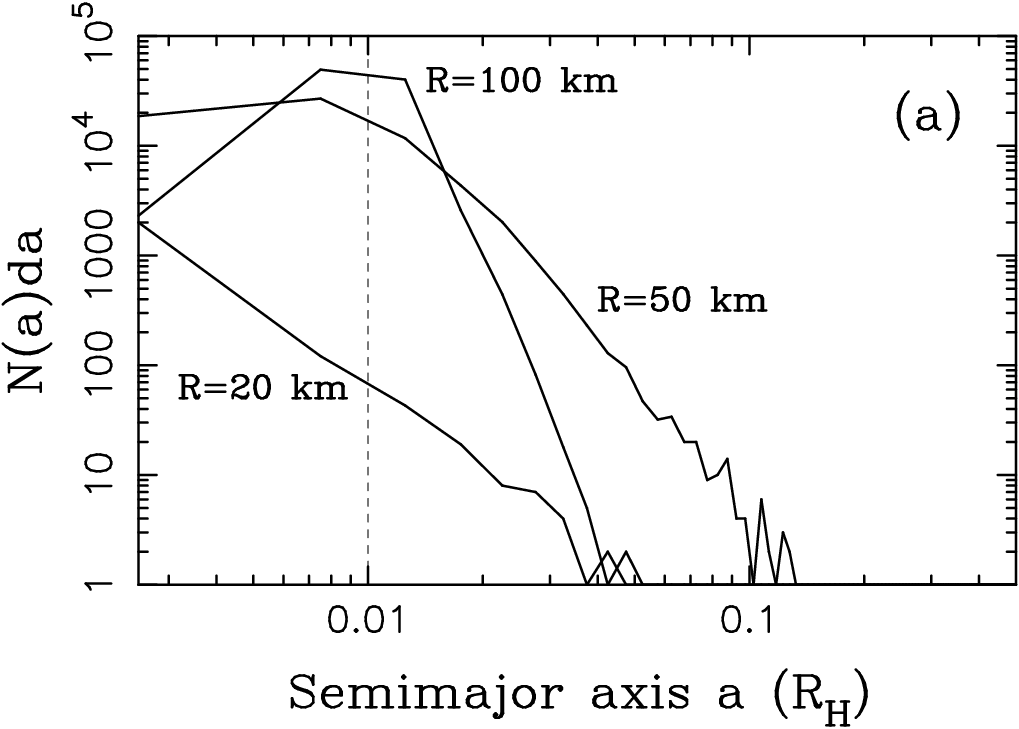}\vspace*{5.mm}
\plotone{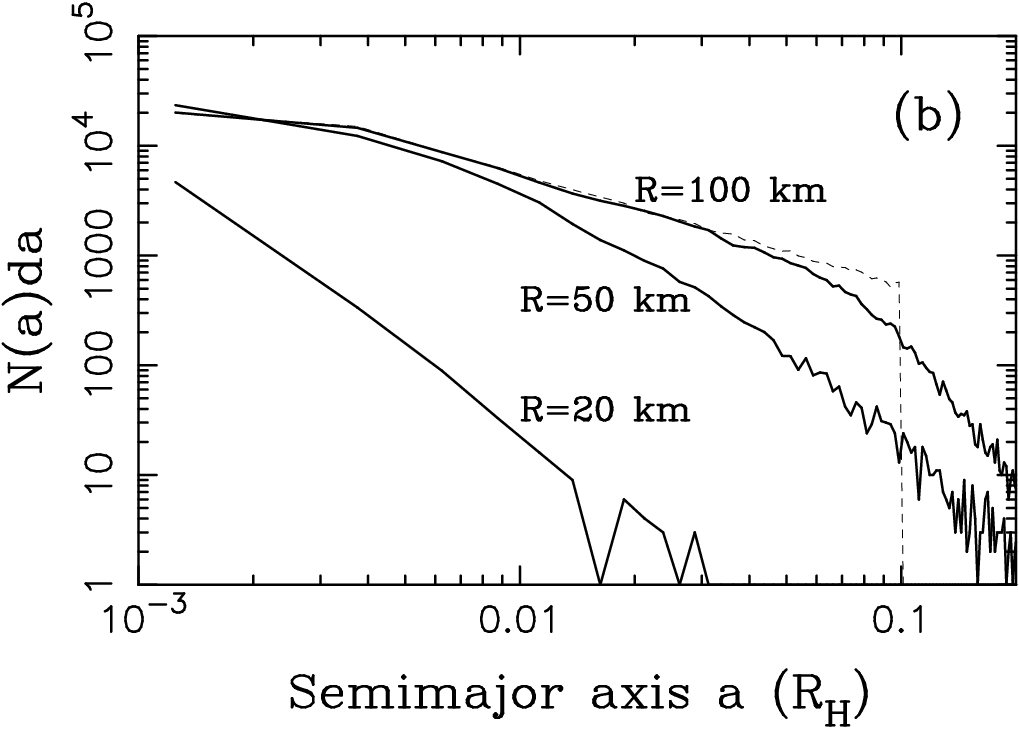}
\caption{The semimajor axis distribution of binary orbits for two different initial distributions. 
(a) All binary orbits have $a=0.01$ $R_{\rm H}$ initially (dashed line). (b) The number of binary orbits per interval 
${\rm d}a$ decreases as $1/a$ for $a<0.1$ $R_{\rm H}$ (dashed line). We plot results for $R=20$, 50 and 
100 km for the Fraser's case illustrated in Fig. \ref{fraser}.}
\label{dsema}
\end{figure}

So far we discussed the binary survival in different collisional scenarios. Here we describe the 
distribution of binary orbits of the surviving binaries. Figure \ref{dsema}a shows the semimajor axis 
distribution of binary orbits produced in simulations with initial $a=0.01$ $R_{\rm H}$, and 
$R=20$, 50 and 100 km. Figure \ref{dsema}b shows the same case for $N(a){\rm d}a\propto{\rm d}a/a$ for 
$a<0.1$~$R_{\rm H}$, and $N(a){\rm d}a=0$ for $a>0.1$ $R_{\rm H}$. These results were obtained for 
the Fraser's case (see Fig.~\ref{fraser}). 

While the binaries with $R\gtrsim100$ km tend to retain the shape of their original distribution, the 
binary orbits for $R\lesssim50$ km become significantly modified. For example, Fig.~\ref{dsema}a shows
that binaries with $R=50$ km and $a=0.01$ $R_{\rm H}$ become tighter or looser, producing 
a wide range of separations with a tail extending to $a=0.1$ $R_{\rm H}$. The number of very wide binaries 
produced by collisions, however, is not expected to be large, because the tail of the distribution at 
$a\sim0.1$ $R_{\rm H}$ in Fig. \ref{dsema}a only represents a very small fraction.

Figure \ref{dsema}b shows how the initial gradient of binary fraction with $a$ can be modified by 
collisions. For example, the initial gradient $\propto a^{-1}$ for $R=20$ km binaries, becomes 
$\propto a^{-3.5}$. Thus, if the collisional effects on 
these binary systems were important, it could be difficult to try to infer their primordial semimajor 
distribution from present observations. On the other hand, the semimajor axis distribution of 
binaries with $R\gtrsim100$ km should not have changed much since their formation. 
 
\section{Comparison with Known Binaries}

All simulations that we conducted so far showed the following result. If the parameters were set up
so that collisional grinding produced the SFD roll-over of cold CKB at $R^*=25$-50 km, the final 
binary fraction showed a strong gradient with radius. Therefore, such a gradient, if identified by 
observations of binary KBOs, would be a direct evidence for the extensive collisional grinding in the 
Kuiper belt. The absence of such a gradient, on the other hand, would indicate that the roll-over is 
more likely accretional.

\begin{figure}
\epsscale{0.6}
\plotone{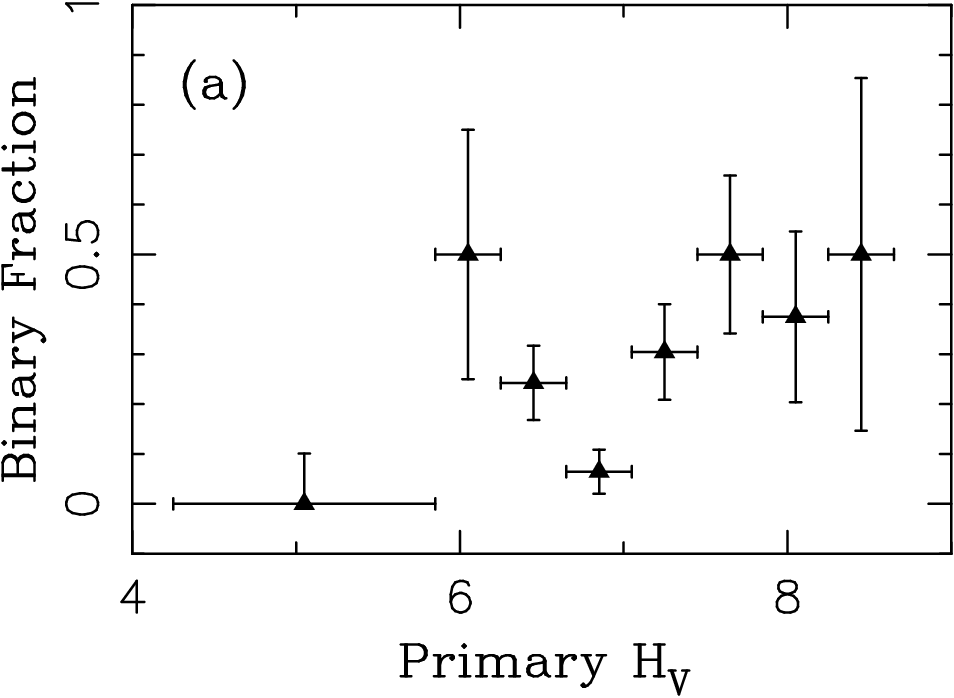}\vspace*{5.mm}
\plotone{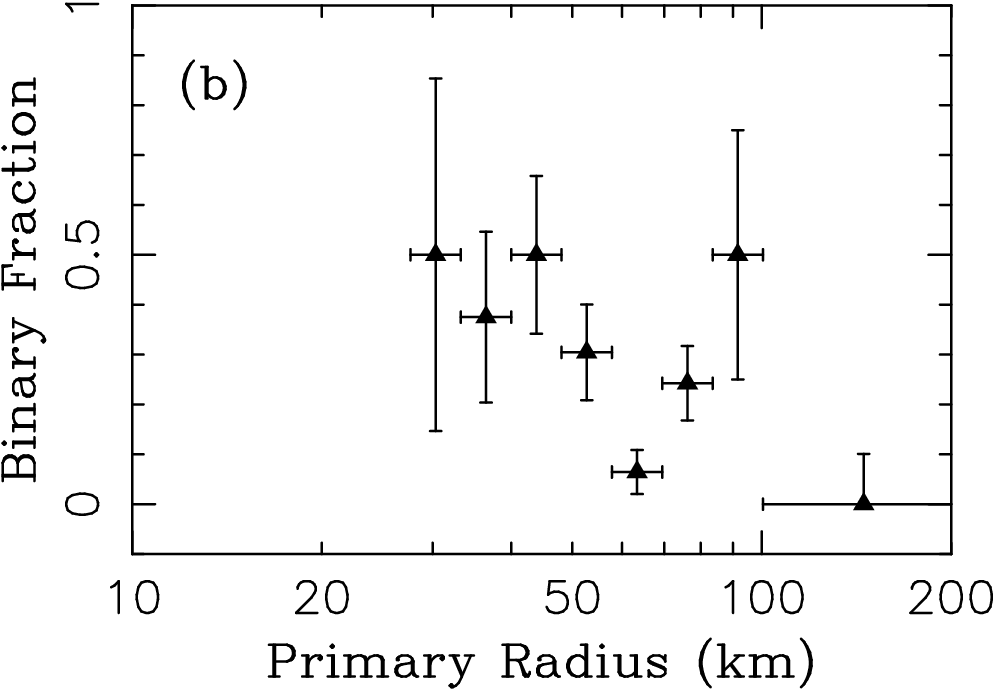}
\caption{The binary fraction in the cold CKB as a function of: (a) absolute primary magnitude 
$H_V$, and (b) primary radius. Data for 122 KBOs with $i<10^\circ$, including 33 binaries, were 
taken from the observational surveys described in Noll et al. (2008a,b). The binary statistics 
for $i<5^\circ$ is very similar to the one shown here. Crudely debiased data (see main text) were 
binned using $\Delta H_V = 0.4$. For $H_V < 5.85$, we combined all data into a single bin. The 
radii of objects in (b) were computed with $p_V=0.2$. The error bars and upper limits of 
non-detections show formal 1-sigma uncertainties.} 
\label{noll}
\end{figure}

Figure \ref{noll} shows the fraction of binaries in the cold CKB that can be inferred from the 
existing observations (Noll et al. 2008a,b). The fraction shown here was roughly corrected for 
the main observational biases. For example, the secondaries can be more difficult to detect near 
small primaries due to their intrinsic faintness, and also because physically smaller components 
are expected to have smaller separations. We normalized each binary to $H_V=8.3$ by scaling 
its component radii and separation by a factor. The binaries with normalized separations smaller
than 0.032 arcsec, which was the smallest separation detected in the dataset, were removed. In total,
only 5 out of 33 binaries were removed by this procedure indicating that the observational bias 
is not overly important. 

The statistical errors shown in Fig. \ref{noll} are large, making it difficult to reach 
definitive conclusions. Still, some interesting features can be pointed out from data.
First of all, the binary fraction for $H_V\lesssim5.5$ ($R\gtrsim100$ km) seems to be lower than the 
one for $H_V>5.5$. Since these large binaries would be relatively resistant to collisions during the 
solar system history, they paucity probably tells us something about the binary formation process.

The fraction of binaries with $H_V>5.5$ ($R<100$ km) does not show any strong gradient with $H_V$ 
(or $R$). Instead, the binary fraction is relatively constant and large down to the smallest surveyed 
objects ($R\sim30$ km). In contrast, we found in \S3 that the effects of collisional grinding 
should deplete binaries with $R\sim30$ km, relative to those with $R\sim100$ km, by a factor of 
$\sim$10. We therefore believe that the existing data are not suggestive of the kind of trends that 
we would expect to see in a population that experienced strong collisional grinding. This may indicate 
that the SFD roll-over in the cold CKB at $R^*=25$-50 km was not produced by disruptive collisions, 
but was instead already in place when KBOs were forming. 

Better observational statistics will be needed, especially for $R\lesssim30$ km, to test this 
preliminary conclusion. Additional caution needs to be exercised when comparing our collisional 
model results with observations, if the formation mechanism was capable of producing the binary 
fraction that strongly varied with $R$. For example, a relatively constant binary fraction could 
result from a combination of the formation and evolution effects, if the initial fraction was 
larger for smaller $R$ and was modified by the collisional removal of small binaries. On the other 
hand, it is unlikely that the initial binary fraction was exactly 100\%, as assumed here. If it 
were lower, our results could be used to place even harder limits on the extent of collisional 
grinding in KB.

Another interesting feature in Fig. \ref{noll} is the dip at $H_V=6.8$ ($R=60$-70 km), where only 2 
out of 31 surveyed cold CKBOs ($\approx$6\%) turned out to be binary. Our collisional 
simulations were capable of producing such a dip (see, e.g. Fig. 4b), but at smaller radii ($R=10$-20 
km). We performed a Monte Carlo search in parameter space to identify cases that could produce 
the observed dip at $R=60$-70 km. The best results were obtained with the initial break at $R>10$ km, 
$v_i<100$ m s$^{-1}$ and substantial initial mass. The low collisional speeds were required here so that the 
dip radius ended up to be only a factor of $\sim$2 larger than the final SFD roll-over radius. The SFD 
for $R>10$ km did not change much in these simulations implying that the SFD roll-over at $R^*=25$-50 
km would have to be pretty much in place before the collisional evolution started. 

\section{Conclusions} 

The work presented here shows how the Kuiper belt could have been affected by collisions.
We found that extensive collisional grinding, required if the SFD roll-over at $R^*=25$-50 km 
in cold CKB were collisional, should imply a strong gradient of binary fraction as a function of $R$ 
and separation. The current observational data do not show signs of such a gradient and instead 
suggest that small binaries ($R<50$ km) are at least as common as the large ones ($R>50$ km). This
may indicate that the SFD roll-over of cold CKBOs at $R^*=25$-50 km is {\it not} due to a prolonged 
phase of collisional grinding in the Kuiper belt. Instead, the roll-over may be a fossil remnant of the 
KBO formation process. Future surveys of small binary KBOs will be able to test this conclusion.       
 
\acknowledgements

We thank P. Benavidez, A. Morbidelli and A. Youdin for useful discussions, and the anonymous reviewer
for helpful suggestions. DN's work was supported 
by the NASA OPR program. The work of DV was partially supported by the Czech Grant Agency (grant 
205/08/0064) and the Research Program MSM0021620860 of the Czech Ministry of Education.

\clearpage

\clearpage

\clearpage

\clearpage

\clearpage

\clearpage

\clearpage

\clearpage

\clearpage

\clearpage


\begin{thebibliography}{}

\bibitem[Astakhov et al.(2005)]{2005MNRAS.360..401A} Astakhov, S.~A., Lee, 
E.~A., \& Farrelly, D.\ 2005, MNRAS, 360, 401

\bibitem{ben} Benavidez, P. G., Campo Bagatin, A. C. 2009. Planetary and Space Science, 
57, 201

\bibitem{ben2} Benavidez, P. G., et al. 2011. In preparation.

\bibitem[Benz 
\& Asphaug(1999)]{1999Icar..142....5B} Benz, W., \& Asphaug, E.\ 1999, Icarus, 142, 5 

\bibitem[Bernstein et al.(2004)]{2004AJ....128.1364B} Bernstein, G.~M., 
Trilling, D.~E., Allen, R.~L., Brown, M.~E., Holman, M., \& Malhotra, R.\ 
2004, AJ, 128, 1364 

\bibitem{ber} Bertotti, B., Farinella, P., \& Vokrouhlick\'y, D. 2003, Physics of 
the Solar System, Kluwer

\bibitem[Bottke et al.(2005)]{2005Icar..175..111B} Bottke, W.~F., Durda, 
D.~D., Nesvorn{\'y}, D., Jedicke, R., Morbidelli, A., Vokrouhlick{\'y}, D., 
\& Levison, H.\ 2005, Icarus, 175, 111 

\bibitem[Bottke et al.(2010)]{2010AJ....139..994B} Bottke, W.~F., 
Nesvorn{\'y}, D., Vokrouhlick{\'y}, D., 
\& Morbidelli, A.\ 2010, \aj, 139, 994 

\bibitem[Brown et al.(2007)]{2007Natur.446..294B} Brown, M.~E., Barkume, 
K.~M., Ragozzine, D., \& Schaller, E.~L.\ 2007, \nat, 446, 294 

\bibitem[Brucker et al.(2009)]{2009Icar..201..284B} Brucker, M.~J., Grundy, 
W.~M., Stansberry, J.~A., Spencer, J.~R., Sheppard, S.~S., Chiang, E.~I., 
\& Buie, M.~W.\ 2009, Icarus, 201, 284 

\bibitem[Chiang 
\& Youdin(2010)]{2010AREPS..38..493C} Chiang, E., \& Youdin, A.~N.\ 2010, Annual 
Review of Earth and Planetary Sciences, 38, 493 

\bibitem[Davis 
\& Farinella(1997)]{1997Icar..125...50D} Davis, D.~R., \& Farinella, P.\ 1997, Icarus, 125, 50 

\bibitem[Davis et al.(2002)]{2002aste.conf..545D} Davis, D.~R., Durda, 
D.~D., Marzari, F., Campo Bagatin, A., 
\& Gil-Hutton, R.\ 2002, Asteroids III, 545 

\bibitem[Dell'Oro 
\& Cellino(2007)]{2007MNRAS.380..399D} Dell'Oro, A., \& Cellino, A.\ 2007, \mnras, 380, 399 

\bibitem[Dell'Oro et 
al.(2001)]{2001A&A...366.1053D} Dell'Oro, A., Marzari, F., Paolicchi, P., \& Vanzani, V.\ 2001, \aap, 366, 1053 

\bibitem[Dohnanyi(1969)]{1969JGR....74.2531D} Dohnanyi, J.~S.\ 1969, \jgr, 
74, 2531 

\bibitem[Dones et al.(2004)]{2004come.book..153D} Dones, L., Weissman, 
P.~R., Levison, H.~F., \& Duncan, M.~J.\ 2004, Comets II, 153 

\bibitem[Donnison(2010)]{2010P&SS...58.1169D} Donnison, J.~R.\ 2010, \planss, 58, 1169 

\bibitem[Durda et al.(2004)]{2004Icar..170..243D} Durda, D.~D., Bottke, 
W.~F., Enke, B.~L., Merline, W.~J., Asphaug, E., Richardson, D.~C., 
\& Leinhardt, Z.~M.\ 2004, Icarus, 170, 243 

\bibitem[Durda et al.(2007)]{2007Icar..186..498D} Durda, D.~D., Bottke, 
W.~F., Nesvorn{\'y}, D., Enke, B.~L., Merline, W.~J., Asphaug, E., 
\& Richardson, D.~C.\ 2007, Icarus, 186, 498 

\bibitem[Fraser(2009)]{2009ApJ...706..119F} Fraser, W.~C.\ 2009, \apj, 706, 
119 

\bibitem{fra1} Fraser, W. C., Brown, M. E., \& Schwamb, M. E. 2010, Icarus, in 
press

\bibitem[Fuentes 
\& Holman(2008)]{2008AJ....136...83F} Fuentes, C.~I., \& Holman, M.~J.\ 2008, \aj, 136, 83 

\bibitem[Fuentes et al.(2009)]{2009ApJ...696...91F} Fuentes, C.~I., George, 
M.~R., \& Holman, M.~J.\ 2009, \apj, 696, 91 

\bibitem[Funato et al.(2004)]{2004Natur.427..518F} Funato, Y., Makino, J., 
Hut, P., Kokubo, E., \& Kinoshita, D.\ 2004, Nature, 427, 518

\bibitem[Gladman et al.(2008)]{2008ssbn.book...43G} Gladman, B., Marsden, 
B.~G., \& Vanlaerhoven, C.\ 2008, The Solar System Beyond Neptune, 43 

\bibitem[Goldreich et al.(2002)]{2002Natur.420..643G} Goldreich, P., 
Lithwick, Y., \& Sari, R.\ 2002, Nature, 420, 643 

\bibitem[Gomes et al.(2005)]{2005Natur.435..466G} Gomes, R., Levison, 
H.~F., Tsiganis, K., \& Morbidelli, A.\ 2005, \nat, 435, 466 

\bibitem[Grundy et al.(2009)]{2009Icar..200..627G} Grundy, W.~M., Noll, 
K.~S., Buie, M.~W., Benecchi, S.~D., Stephens, D.~C., \& Levison, H.~F.\ 
2009, Icarus, 200, 627

\bibitem[Jewitt 
\& Luu(1995)]{1995AJ....109.1867J} Jewitt, D.~C., \& Luu, J.~X.\ 1995, \aj, 109, 1867 

\bibitem[Kenyon \& Bromley(2001)]{2001AJ....121..538K} Kenyon, S.~J., \& Bromley, 
B.~C.\ 2001, \aj, 121, 538 

\bibitem[Kenyon et al.(2008)]{2008ssbn.book..293K} Kenyon, S.~J., Bromley, 
B.~C., O'Brien, D.~P., 
\& Davis, D.~R.\ 2008, The Solar System Beyond Neptune, 293 

\bibitem[Kozai(1962)]{1962AJ.....67..591K} Kozai, Y.\ 1962, \aj, 67, 591 

\bibitem[Leinhardt and Stewart(2009)]{2009Icar..199..542L} Leinhardt, 
Z.~M., \& Stewart, S.~T.\ 2009.\ Icarus, 199, 542

\bibitem[Levison 
\& Duncan(1997)]{1997Icar..127...13L} Levison, H.~F., \& Duncan, M.~J.\ 1997, Icarus, 127, 13 

\bibitem[Levison et al.(2008)]{2008Icar..196..258L} Levison, H.~F., 
Morbidelli, A., Vanlaerhoven, C., Gomes, R., 
\& Tsiganis, K.\ 2008a, Icarus, 196, 258 

\bibitem[Levison et al.(2008)]{2008AJ....136.1079L} Levison, H.~F., 
Morbidelli, A., Vokrouhlick{\'y}, D., 
\& Bottke, W.~F.\ 2008b, \aj, 136, 1079 

\bibitem[Levison et al.(2009)]{2009Natur.460..364L} Levison, H.~F., Bottke, 
W.~F., Gounelle, M., Morbidelli, A., Nesvorn{\'y}, D., 
\& Tsiganis, K.\ 2009, \nat, 460, 364 

\bibitem[Merline et al.(2002)]{2002aste.conf..289M} Merline, W.~J., 
Weidenschilling, S.~J., Durda, D.~D., Margot, J.~L., Pravec, P., 
\& Storrs, A.~D.\ 2002, Asteroids III, 289 

\bibitem[Morbidelli et al.(2008)]{2008ssbn.book..275M} Morbidelli, A., 
Levison, H.~F., \& Gomes, R.\ 2008, The Solar System Beyond Neptune, 275 

\bibitem[Morbidelli et al.(2009)]{2009Icar..204..558M} Morbidelli, A., 
Bottke, W.~F., Nesvorn{\'y}, D., \& Levison, H.~F.\ 2009, Icarus, 204, 558 

\bibitem[Nesvorn{\'y} et al.(2003)]{2003AJ....126..398N} Nesvorn{\'y}, D., 
Alvarellos, J.~L.~A., Dones, L., \& Levison, H.~F.\ 2003, \aj, 126, 398 

\bibitem[Nesvorn{\'y} et al.(2006)]{2006Icar..183..296N} Nesvorn{\'y}, D., 
Enke, B.~L., Bottke, W.~F., Durda, D.~D., Asphaug, E., 
\& Richardson, D.~C.\ 2006, Icarus, 183, 296 

\bibitem[Nesvorn{\'y} et al.(2010)]{2010AJ....140..785N} Nesvorn{\'y}, D., 
Youdin, A.~N., \& Richardson, D.~C.\ 2010, \aj, 140, 785 

\bibitem[Noll et al.(2008)]{2008ssbn.book..345N} Noll, K.~S., Grundy, 
W.~M., Chiang, E.~I., Margot, J.-L., \& Kern, S.~D.\ 2008a, 
In The Solar System Beyond Neptune, 345 

\bibitem[Noll et al.(2008)]{2008Icar..194..758N} Noll, K.~S., Grundy, 
W.~M., Stephens, D.~C., Levison, H.~F., \& Kern, S.~D.\ 2008b, 
Icarus, 194, 758

\bibitem[O'Brien 
\& Greenberg(2003)]{2003Icar..164..334O} O'Brien, D.~P., \& Greenberg, R.\ 2003, Icarus, 164, 334 

\bibitem[Pan 
\& Sari(2005)]{2005Icar..173..342P} Pan, M., \& Sari, R.\ 2005, Icarus, 173, 342 

\bibitem[Parker 
\& Kavelaars(2010)]{2010ApJ...722L.204P} Parker, A.~H., \& Kavelaars, J.~J.\ 2010, \apjl, 722, L204 

\bibitem[Petit and Mousis(2004)]{2004Icar..168..409P} Petit, J.-M., \& Mousis, 
O.\ 2004, Icarus, 168, 409

\bibitem[Petit et al.(2008)]{2008ssbn.book...71P} Petit, J.-M., Kavelaars, 
J.~J., Gladman, B., 
\& Loredo, T.\ 2008, The Solar System Beyond Neptune, 71 

\bibitem{sch} Schaller, E. L., TNO 2010 meeting, Dynamical and Physical Properties of 
Trans-Neptunian Objects, Philadelphia  

\bibitem[Schlichting et al.(2009)]{2009Natur.462..895S} Schlichting, H.~E., 
Ofek, E.~O., Wenz, M., Sari, R., Gal-Yam, A., Livio, M., Nelan, E., 
\& Zucker, S.\ 2009, \nat, 462, 895 

\bibitem[Stern and Colwell(1997)]{1997AJ....114..841S} Stern, S.~A., 
\& Colwell, J.~E.\ 1997, AJ, 114, 841

\bibitem[Stern et al.(2003)]{2003AJ....125..902S} Stern, S.~A., Bottke, 
W.~F., \& Levison, H.~F.\ 2003, \aj, 125, 902 

\bibitem[Stewart and Leinhardt(2009)]{2009ApJ...691L.133S} Stewart, S.~T., 
\& Leinhardt, Z.~M.\ 2009, ApJ, 691, L133 

\bibitem[Tegler 
\& Romanishin(2000)]{2000Natur.407..979T} Tegler, S.~C., \& Romanishin, W.\ 2000, \nat, 407, 979 

\bibitem{tru} Trujillo, C.~A., Jewitt, D. C., \& Luu, J. X. 2001, AJ, 122, 457

\bibitem[Tsiganis et al.(2005)]{2005Natur.435..459T} Tsiganis, K., Gomes, 
R., Morbidelli, A., \& Levison, H.~F.\ 2005, \nat, 435, 459 

\bibitem[Weidenschilling(2002)]{2002Icar..160..212W} Weidenschilling, 
S.~J.\ 2002, Icarus, 160, 212 

\bibitem[Weidenschilling et al.(1997)]{1997Icar..128..429W} 
Weidenschilling, S.~J., Spaute, D., Davis, D.~R., Marzari, F., 
\& Ohtsuki, K.\ 1997, Icarus, 128, 429 

\end{thebibliography}
\end{document}